\documentclass[pra,twocolumn,floatfix]{revtex4}
\usepackage{graphicx}
\usepackage{amsmath}
\usepackage{amssymb}
\usepackage{hyperref}
\hypersetup{colorlinks=true, linkcolor=blue, citecolor=blue, urlcolor=blue}

\newcommand{\ket}[1]{\left| #1 \right>} 
\newcommand{\bra}[1]{\left< #1 \right|} 

\begin{document}

\title{Synthetic spin-orbit interactions and magnetic fields in ring-cavity QED}
\author{Farokh Mivehvar}
\author{David L. Feder}
\email[Corresponding author: ]{dfeder@ucalgary.ca}
\affiliation{Institute for Quantum Science and Technology, 
Department of Physics and Astronomy,
University of Calgary, Calgary, Alberta, Canada T2N 1N4}
\date{\today}
\begin{abstract}
The interactions between light and matter are strongly enhanced when atoms are
placed in high-finesse quantum cavities, offering tantalizing opportunities for
generating exotic new quantum phases. In this work we show that both 
spin-orbit interactions and strong synthetic magnetic fields result when 
a neutral atom is confined within a ring cavity, whenever the internal atomic 
states are coupled to two off-resonant counter-propagating modes. We 
diagonalize the resulting cavity polariton Hamiltonian and find characteristic 
spin-orbit dispersion relations for a wide range of parameters. An adjustable 
uniform gauge potential is also generated, which can be converted into a
synthetic magnetic field for neutral atoms by applying an external magnetic
field gradient. Very large synthetic magnetic fields are possible as the 
strength is proportional to the (average) number of photons in each of the 
cavity modes. The results suggest that strong-coupling cavity quantum 
electrodynamics can be a useful environment for the formation of topological 
states in atomic systems.
\end{abstract}

\maketitle

%
%

\section{Introduction}
\label{sec:introduction}

The spin-orbit interaction in solids is the coupling of an electron's spin to 
its center-of-mass momentum, and is closely related to the spin-orbit coupling
in atomic systems. In two-dimensional electron gases two kinds of spin-orbit
coupling have important effects on the electronic band structure: 
Dresselhaus~\cite{Dresselhaus-1955} and Rashba~\cite{Rashba-1984} interactions. 
In a groundbreaking paper~\cite{Kane-2005-a}, Kane and Mele showed that 
including a spin-orbit interaction in the Hamiltonian of graphene, while 
respecting all of the material's symmetries, nevertheless opens up a band gap. 
The resulting bands become topologically nontrivial, so that the material 
supports a pair of robust conducting edge states characterized by a nontrivial 
$Z_2$ 
topological invariant~\cite{Kane-2005-b}. This new phase of matter is known as
a topological insulator, or a quantum spin Hall (QSH) insulator in two 
dimensions, and its discovery has opened up a fascinating new research area in 
condensed matter physics~\cite{Hasan-2010}. Determining the conditions under
which topological states could arise in condensed matter systems is the subject 
of continuing investigations~\cite{Schnyder-2008,Wen-2012,Slager-2013}.

Ultracold atomic gases provide a rich environment for the simulation of 
condensed matter physics~\cite{Jaksch-2005,Lewenstein-2007,Bloch-2008}. For
example, interacting atoms confined in optical lattices experience a 
crystalline environment that can mimic strongly correlated superfluid and 
magnetic states. Over the past decade, many theoretical schemes have been 
proposed to generate synthetic gauge potentials for neutral ultracold atomic 
gases via atom-light interactions~\cite{Dalibard-2011}. In recent years, both
synthetic magnetic~\cite{Lin-2009} and electric~\cite{Lin-2011-b} fields have
been realized experimentally. The spin-orbit coupling can be interpreted as a 
non-Abelian gauge field~\cite{Hatano-2007}, and Lin~{\it et al.} recently 
realized a scheme to generate a combination of Rashba and Dresselhaus 
spin-orbit couplings in ultracold neutral atoms by means of resonant two-photon
Raman 
transitions~\cite{Lin-2011-a}. The strength of the gauge field potentials in
these experiments is limited by the atomic recoil momentum, though there are
recent theoretical proposals that would push these to much higher 
values~\cite{Cooper-2012-a,Cooper-2012-b,Cooper-2013}.

Placing atoms in high-finesse optical cavities strongly enhances atom-photon
interactions~\cite{Walther-2006}, with numerous potential applications to 
quantum information science~\cite{Raimond-2008}. While much of the early work
focused on single atoms, recent investigations of cavity quantum 
electrodynamics (QED) with multiple trapped ultracold atoms~\cite{Colombe-2007}
are revealing fascinating new phenomena. The field mode 
to which atoms are collectively coupled is in turn affected by the atomic 
states, giving rise to cavity mediated long-range atom-atom 
interactions~\cite{Ritsch-2012}. Other examples include the Dicke phase 
transition~\cite{Baumann-2010} and a collective atomic recoil 
laser~\cite{Kruse-2003,Cube-2004,Slama-2007,Slama-2007-A} in many-atom linear 
and ring cavity QED, respectively.

The strong coupling of cavity QED therefore offers the tantalizing prospect of 
enhancing the magnitude of synthetic gauge fields and spin-orbit interactions 
in atomic systems, as well as inducing unique strongly correlated states of 
both atoms and photons with no analog in condensed matter systems. In this 
work, we show how to simultaneously engineer a spin-orbit interaction and a 
synthetic magnetic field 
for a single neutral atom confined inside a ring cavity, as a first step toward
generating topological states in ultracold atomic systems. We build on the 
central ideas of two-photon resonant Raman transitions described in 
Ref.~\cite{Lin-2011-a}, in which absorption and re-emission of photons from one
beam to the other naturally couples the atom's internal states to their 
center-of-mass
momentum. Two propagating modes of a high-finesse ring cavity accomplish the 
same purpose, but with an enhanced atom-photon coupling strength. This 
potential advantage comes at the cost of increased mathematical complexity,
because unlike the continuum Raman case both the atom and photon degrees of 
freedom need to be treated fully quantum mechanically. 

The calculations presented here reveal that the spin-orbit interactions and
synthetic magnetic fields emerge naturally as the limits of zero two-photon 
detuning between the atomic and cavity frequencies and large two-photon 
detuning, respectively. The spin-orbit interactions are only weakly dependent 
on the occupation of the 
cavity modes, and in fact are robust already at the level of a few photons.
That said, the energy barrier between the energy levels split by the spin-orbit
interactions is greatest when the difference between the occupation of the two
modes is largest. In principle, this parameter is adjustable 
experimentally~\cite{Brattke-2001,McKeever-2004,Keller-2004,Merlin-Cooper-2013}.
The strength of the synthetic magnetic 
fields is proportional to the square of the total number of photons in the 
cavity. The cavity QED environment therefore promises huge synthetic magnetic 
fields, potentially much larger than are currently accessible to ultracold atom 
experiments. The readiness with which spin-orbit interactions and synthetic 
magnetic fields are manifested in cavity QED should facilitate the production 
of new strongly correlated states in these systems.

The manuscript is organized as follows. The model of the atom interacting with
a ring cavity is described in Sec.~\ref{sec:model}, and the governing 
Hamiltonian is derived. In Sec.~\ref{sec:polaritons}, this Hamiltonian is 
expressed in terms of polaritons and diagonialized to obtain the spectrum of 
excitations. Sec.~\ref{sec:syntheticfields} describes the circumstances under 
which synthetic spin-orbit interactions and magnetic fields emerge in this 
model. Sec.~\ref{sec:conclusions} discusses the results with a view toward 
future calculations.

%
%

\section{Model and Hamiltonian}
\label{sec:model}

\begin{figure}
\centering
\includegraphics [width=0.4\textwidth]{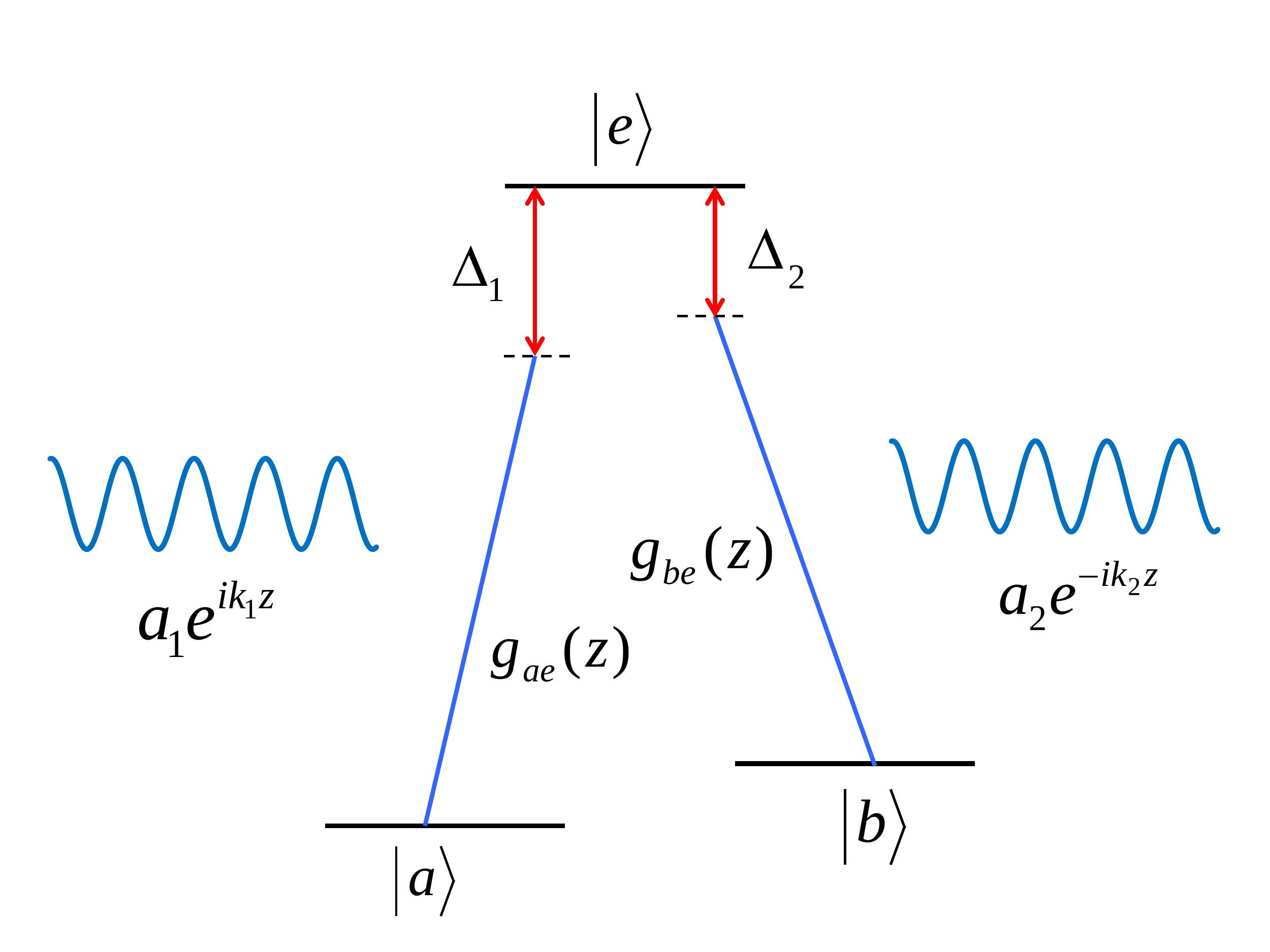}
\caption{Two low-lying atomic levels $\ket{a}$ and $\ket{b}$ are coupled in
the $\Lambda$ scheme to an excited state $\ket{e}$ by two counter-propagating 
cavity modes $a_1e^{\imath k_1z}$ and $a_2e^{-\imath k_2z}$ with strength 
$g_{ae}(z)$ and $g_{be}(z)$, respectively. The transitions are detuned from 
resonance by $\Delta_1$ and $\Delta_2$.}
	\label{fig:Lambda-scheme}
\end{figure}

Consider a ring cavity with two counter-propagating modes $a_1e^{\imath k_1z}$ 
and $a_2e^{-\imath k_2z}$, where $a_i$ are field annihilation operators for 
the photon and $k_i=\omega_i/c$ are the photon wavenumber expressed in terms 
of their frequencies $\omega_i$. Note that in this work $\imath\equiv\sqrt{-1}$.
Three atomic levels are coupled via these two cavity 
modes in the $\Lambda$ scheme, as depicted in Fig.~\ref{fig:Lambda-scheme}. 
The states $\ket{a}$, $\ket{b}$, and $\ket{e}$ are arbitrary internal states
of an atom whose energies satisfy the relations $E_e>E_b>E_a$ and 
$\{E_{ea},E_{eb}\}\gg E_{ba}$, where $E_{ij}\equiv E_i-E_j$. For example, the 
states $\ket{a}$ and 
$\ket{b}$ might be energy levels in the same hyperfine manifold with an
energy separation on the order of MHz while the state $\ket{e}$ could be an 
excited electronic level with an energy separation on the order of THz. The
mode $a_1e^{\imath k_1z}$ ($a_2e^{-\imath k_2z}$) propagates to the right 
(left) and couples solely to the $\ket{a}\leftrightarrow\ket{e}$ 
($\ket{b}\leftrightarrow\ket{e}$) transition. 

The Hamiltonian in the rotating-wave approximation reads
\begin{eqnarray} 
H&=&\frac{\hbar^2q_z^2}{2m}I_{3 \times 3}+E_a \sigma_{aa}+E_b \sigma_{bb}
+E_e \sigma_{ee}\nonumber \\
&+&\hbar(\omega_1a^\dagger_1a_1+\omega_2a_2^\dagger a_2)
\nonumber \\ 
&+&\hbar\left( g_{ae}(z)a_1\sigma_{ea}+g_{be}(z)a_2\sigma_{eb}+\mbox{H.c.}
\right),
\label{eq:Hamiltonian-RWA}
\end{eqnarray}
where $\sigma_{ij}=\ket{i}\bra{j}$, $g_{ae}(z)=g_{ae}e^{\imath k_1z}$, 
$g_{be}(z)=g_{be}e^{-\imath k_2z}$ and H.c. stands for Hermitian conjugate. 
Here $\hbar q_z$ is the center-of-mass momentum of the atom and 
$I_{3 \times 3}$ is the identity matrix in the internal atomic state space. 
Note that strictly speaking, Eq.~(\ref{eq:Hamiltonian-RWA}) is the Hamiltonian
density; alternatively it can be considered the Hamiltonian with an implied sum 
over $q_z$. To a first approximation, this Hamiltonian represents an 
atom with infinite excited-state lifetime and an exceptionally high-finesse 
cavity, neglecting atomic spontaneous emission and 
cavity losses by mirror leakage, as well as gains by external pumps.
Even with these simplifying assumptions, the analysis of this Hamiltonian is 
quite involved as can be seen below; relaxing these assumptions will therefore 
be the focus of future work.

If the cavity modes are far off-resonance from atomic transitions 
then the adiabatic condition holds. That is, if the frequency detunings 
$\hbar\Delta_1\equiv\hbar\omega_1-E_{ea}$
and $\hbar\Delta_2\equiv\hbar\omega_2-E_{eb}$ are very large compared to 
$E_{ba}$ then the excited state $\ket{e}$ can be adiabatically 
eliminated~\cite{Gerry-1990}. Details of the procedure can be found in the
Appendix~\ref{sec:Adiabatic-Elimination}. The effective Hamiltonian for the 
ground pseudospin states $\ket{a}$ and $\ket{b}$ and the cavity modes then 
becomes
\begin{eqnarray} \label{eq:Hamiltonian-adiabatic}
H_{\rm eff}&=&\frac{\hbar^2q_z^2}{2m}I_{2\times 2}
+\frac{1}{2}\hbar\tilde{\omega}_0\sigma_z
+\hbar(\omega_1a^\dagger_1a_1+\omega_2a_2^\dagger a_2)\nonumber\\
&+&\hbar\Omega_R\left(e^{\imath (k_1+k_2)z}a_2^\dagger a_1\sigma_{ba}
+\mbox{H.c.}\right),
\end{eqnarray}
where $\sigma_{z}=|b\rangle\langle b|-|a\rangle\langle a|$, and 
$\hbar\tilde{\omega}_0=\tilde{E}_b-\tilde{E}_a>0$. The $\tilde{E}_i$ correspond
to the AC Stark-shifted atomic energies~(\ref{ACStark}). The $I_{2 \times 2}$ 
operator is the identity matrix in the ground pseudospin state space, and will 
be implied in the remainder of this work. The two-photon Rabi 
frequency~(\ref{Rabi}) is given by $\Omega_R=g_{ae}g_{be}\left(
\frac{\Delta_1+\Delta_2}{\Delta_1\Delta_2}\right)$ under the assumption that
$\{g_{ae},g_{be}\}\in\Re$.

It is useful to perform a Galilean transformation of this Hamiltonian into the 
frame moving at the momentum transferred to the atom by the interaction with 
the photons. This is accomplished using the unitary operator
$U=\exp{[\imath (k_1\sigma_{aa}-k_2\sigma_{bb})z]}$:
\begin{eqnarray}
\tilde{H}_{\rm eff}\equiv UH_{\rm eff}U^\dagger
&=&\frac{\hbar^2}{2m}[q_z-(k_1\sigma_{aa}-k_2\sigma_{bb})]^2\nonumber \\
&+&\frac{1}{2}\hbar\tilde{\omega}_0\sigma_z
+\hbar(\omega_1a_1^\dagger a_1+\omega_2a_2^\dagger a_2)\nonumber\\
&+&\hbar\Omega_R\left(a_2^{\dag}a_1\sigma_{ba}+\mbox{H.c.}\right),
\label{eq:Hamiltonian-Unitary}
\end{eqnarray}
where $U\sigma_{ba}U^{\dag}=e^{-\imath (k_1+k_2)z}\sigma_{ba}$
using the Baker-Campbell-Hausdorff formula. 
One could have applied the unitary 
transformation $U'=\exp{[\imath (k_1a_1^\dagger a_1-k_2a_2^\dagger a_2)z]}$ instead; 
although the first term of the resulting Hamiltonian will be different from
that given above, the final results discussed below are independent of the 
particular choice of a unitary transformation. The last term in 
Eq.~(\ref{eq:Hamiltonian-Unitary}) resembles the interaction term in the 
Jaynes-Cummings model \cite{Shore-1993}, but the $a$ is replaced by a two-photon $a_2^{\dag}a_1$ 
operator. 

In order to reveal the underlying symmetries of the 
Hamiltonian~(\ref{eq:Hamiltonian-Unitary}), it is useful to express the 
operators in the Schwinger representation. Let 
$\sigma_+=\sigma_{ba}=\frac{1}{2}(\sigma_x+\imath\sigma_y)=\frac{1}{\hbar}
(s_x+\imath s_y)=\frac{1}{\hbar}s_+$ and $\sigma_-=\sigma_{ab}
=\frac{1}{\hbar}s_-$ be the 
raising and lowering operators for the atom, and $\frac{2}{\hbar}s_z=\sigma_z
=\sigma_+\sigma_- - \sigma_-\sigma_+=|b\rangle\langle b|-|a\rangle\langle a|$.
If there are only exactly two modes of the cavity and $\omega_1>\omega_2$, one
can make use of the Schwinger angular momentum operators~\cite{Schwinger-1965}
for the photon field operators $j_x=\frac{\hbar}{2}(a_1^\dagger a_2+a_2^\dagger a_1)$, $j_y=\frac{\hbar}{2\imath}(a_1^\dagger a_2-a_2^\dagger a_1)$, and
$j_z=\frac{\hbar}{2}(a_1^\dagger a_1-a_2^\dagger a_2)$, which satisfy the 
$SU(2)$ Lie algebra (or angular momentum commutation relation) 
\begin{equation}
[j_n,j_m]=\imath\hbar \varepsilon^{nml} j_l,
\label{eq:su(2)-algebra}
\end{equation}
where $\varepsilon^{nml}$ is the totally antisymmetric tensor. As in the atomic
case, one can define photonic angular-momentum raising and lowering operators
$j_+=j_x+\imath j_y=\hbar a_1^\dagger a_2$ and $j_-=j_x-\imath j_y
=\hbar a_2^\dagger a_1$.
The Hamiltonian~(\ref{eq:Hamiltonian-Unitary}) can then be recast as 
\begin{eqnarray}
\tilde{H}_{\rm eff}&=&\frac{\hbar^2}{2m}\left\{q_zI_{2\times2}-\left[\frac{\Delta k}{2}I_{2\times2}-k\sigma_z\right]\right\}^2\nonumber \\
&+&\tilde{\omega}_0 s_z+\frac{\hbar}{2}(\omega_1+\omega_2)N
+(\omega_1-\omega_2)j_z\nonumber\\
&+&\frac{\Omega_R}{\hbar}(j_-s_+ +j_+s_-),
\label{eq:Hamiltonian-Using-j-s}
\end{eqnarray}
where $k=(k_1+k_2)/2$, $\Delta k=k_1-k_2$ and 
$I_{2\times2}=\sigma_{aa}+\sigma_{bb}$ as before.
Here, $N=a_1^\dagger a_1+a_2^\dagger a_2$ is the total photon number operator 
with eigenvalues $n=2j$, where $\hbar^2 j(j+1)$ is the eigenvalue of the total 
photon spin operator $\mathbf{j}^2$. 

The first term in the Hamiltonian~(\ref{eq:Hamiltonian-Using-j-s}) strongly
resembles spin-orbit coupling, with equal contributions of 
Dresselhaus~\cite{Dresselhaus-1955} and Rashba~\cite{Rashba-1984} terms.
Expanding the quadratic operator provides a cross term 
that explicitly couples the linear momentum to the pseudospin degree of 
freedom. A more formal mapping will be discussed in detail in the next section.

Aside from the first term, the Hamiltonian~(\ref{eq:Hamiltonian-Using-j-s})
corresponds to a generalized Jaynes-Cummings model: 
\begin{eqnarray}
H_{\rm GJC}&=&\tilde{\omega}_0 s_z+\frac{\hbar}{2}(\omega_1+\omega_2)N
+(\omega_1-\omega_2)j_z\nonumber\\
&+&\frac{\Omega_R}{\hbar}(j_-s_+ +j_+s_-).
\end{eqnarray}
The components of $\mathbf{j}$ and $\mathbf{s}$ both satisfy the angular 
momentum commutation relation~(\ref{eq:su(2)-algebra}), so one can define the 
total angular momentum operator $\mathbf{J}=\mathbf{j}+\mathbf{s}$. 
Because $[H_{\rm GJC},N]=[H_{\rm GJC},\mathbf{j}^2]=[H_{\rm GJC},J_z]=0$, it is
conventional to represent the eigenstates of $H_{\rm GJC}$ and 
$\tilde{H}_{\rm eff}$ in a basis labeled by the eigenstates of $s_z$, 
$j_z$, and ${\bf j}$, with eigenvalues $\hbar m_s=\pm\hbar/2$, 
$\hbar m_j \equiv \hbar(n_1-n_2)/2$, and 
$\hbar j=(\hbar/2)(n_1+n_2)=(\hbar/2)n$, respectively. For reasons described in
detail below, it turns out to be more convenient to instead express the basis 
in eigenstates of ${\bf j}$, $s_z$, and $J_z$, where the eigenvalues of the 
last quantity $\hbar m_z=\hbar(m_j+m_s)=(\hbar/2)(n_1-n_2\pm 1)$. Note that 
the only component of the total angular momentum operator ${\bf J}$ that 
commutes with $\tilde{H}_{\rm eff}$ is $J_z$. Thus, the symmetry of the spin 
space of the Hamiltonian is reduced to $U(1)$.

%
%
\section{Polariton Mapping}
\label{sec:polaritons}

While the first part of the effective 
Hamiltonian~(\ref{eq:Hamiltonian-Using-j-s}) indicates that the atoms
experience an effective spin-orbit coupling through their interactions with 
the cavity modes, the remainder corresponds to a generalized Jaynes-Cummings 
model. The natural representation of the quasiparticles in the latter model is
that of cavity polaritons (superpositions of atomic and photonic excitations).
Explicit diagonalization of the full polariton Hamiltonian, performed below, 
shows that in fact it is the dressed pseudospin states that experience the 
spin-orbit interactions and synthetic magnetic fields.


\subsection{Diagonalizing the generalized Jaynes-Cummings Hamiltonian}
\label{subsec:GJC}

\begin{figure}[t]
\centering
\includegraphics [width=0.49\textwidth]{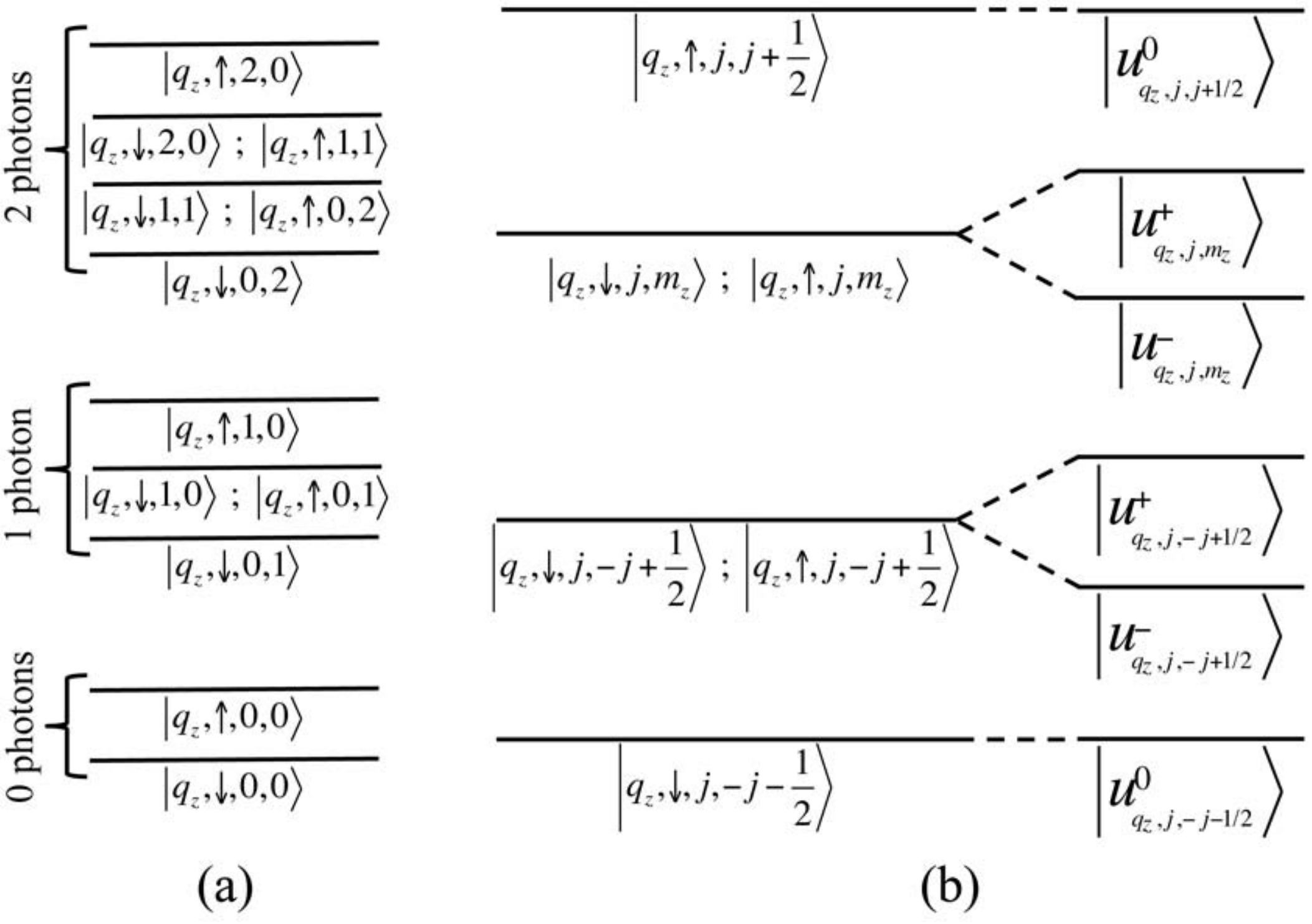}
\caption{The energy manifolds of the atom-cavity system in the uncoupled $q_z$, 
$s_z$, $N_1$, and $N_2$ basis (i.e.\ $\ket{q_z,\uparrow\downarrow,n_1,n_2}$) are
shown in (a). Shown in (b) is the $2j$-photon manifold and the corresponding 
sub-manifolds in the uncoupled $q_z$, $s_z$, $\mathbf{j}$, and $J_z$ basis (i.e. 
$\ket{q_z,\downarrow \uparrow,j,m_z}$), and the resulting dressed 
states of Eq.~(\ref{eq:mixing}) within the manifold. Note that here $\delta=0$.}
\label{fig:energy-manifolds}
\end{figure}

In order to map the Hamiltonian~(\ref{eq:Hamiltonian-Using-j-s}) into the 
polariton basis, we first exactly diagonalize $H_{\rm GJC}$. Ignoring the
atom-photon coupling $j_+s_-+j_-s_+$ term, the natural basis states can be
represented by $\ket{q_z}\otimes\ket{\downarrow\uparrow}\otimes\ket{n_1}\otimes
\ket{n_2}=\ket{q_z,\downarrow\uparrow,n_1,n_2}$, where 
$\ket{\downarrow}=|a\rangle$, $\ket{\uparrow}=|b\rangle$, and $n_1$ and $n_2$
are the number of photons in the first and second cavity modes, respectively. 
With a total of $n$ photons in
the cavity, there are $2(n+1)$ basis states for each value of $q_z$. These
states are depicted in Fig.~\ref{fig:energy-manifolds}. For example, the states 
for $n=1$ correspond to $\ket{q_z,\downarrow,0,1}$, $\ket{q_z,\downarrow,1,0}$, 
$\ket{q_z,\uparrow,0,1}$, and $\ket{q_z,\uparrow,1,0}$. The $j_+s_-+j_-s_+$ 
term couples basis states $\ket{q_z,\downarrow,n_1,n_2}$ and 
$\ket{q_z,\uparrow,n_1-1,n_2+1}$ together within a given $n$ manifold, but the 
states $\ket{q_z,\downarrow,0,n}$ and $\ket{q_z,\uparrow,n,0}$ will remain 
uncoupled. For each value of $n$, the Hamiltonian therefore block diagonalizes 
into $n+2$ distinct blocks, of which $n$ are two-dimensional and two are 
one-dimensional.

Following the discussion in the previous section, it is convenient to represent 
the basis states above in terms of pseudospin quantum numbers:
$\ket{q_z,\downarrow \uparrow,j,m_z}=\ket{q_z}\otimes \ket{\downarrow \uparrow}
\otimes \ket{j,m_z}$, where $\hbar j=\hbar n/2$
and $\hbar m_z=\hbar(m_j+m_s)=\frac{\hbar}{2}(n_1-n_2\pm 1)$ with 
$m_z=-j-\frac{1}{2},\ldots,j+\frac{1}{2}$. In this representation, the states 
$\ket{q_z,\downarrow,j,-j-\frac{1}{2}}\equiv\ket{u^0_{q_z,j,-j-\frac{1}{2}}}$ 
and $\ket{q_z,\uparrow,j,j+\frac{1}{2}}\equiv\ket{u^0_{q_z,j,j+\frac{1}{2}}}$ 
are independent of the others, and have energies 
\begin{eqnarray}
E^0_{j,-j-\frac{1}{2}}&=&-\frac{\hbar\tilde{\omega}_0}{2}+\hbar\omega_2n_2
=-\frac{\hbar\tilde{\omega}_0}{2}+2\hbar\omega_2j;\nonumber \\
E^0_{j,j+\frac{1}{2}}&=&\frac{\hbar\tilde{\omega}_0}{2}+\hbar\omega_1 n_1
=\frac{\hbar\tilde{\omega}_0}{2}+2\hbar\omega_1j. 
\label{E0}
\end{eqnarray}
The remaining $2n$ states couple in pairs keeping $m_z$ fixed. For example, 
states with $m_j=\frac{1}{2}(n_1-n_2)$ and $m_s=-\frac{1}{2}$ (atomic state 
$\ket{\downarrow}=\ket{a}$) couple with states with 
$m_j'=\frac{1}{2}(n_1'-n_2')=\frac{1}{2}[(n_1-1)-(n_2+1)]=m_j-1$ and 
$m_s=+\frac{1}{2}$; both these have $m_z=\frac{1}{2}(n_1-n_2-1)
=m_j-\frac{1}{2}$. The two-dimensional blocks of the 
Hamiltonian~$H_{\rm GJC}$ are therefore
\begin{widetext}
\begin{eqnarray}
H_{\rm GJC}^{j,m_z}&=&
\begin{pmatrix}
\frac{\hbar\tilde{\omega}_0}{2}+\hbar\omega_1(n_1-1)+\hbar\omega_2(n_2+1)
& \hbar\Omega_R\sqrt{n_1\left(n_2+1\right)} \cr
\hbar\Omega_R\sqrt{n_1\left(n_2+1\right)} &
-\frac{\hbar\tilde{\omega}_0}{2}+\hbar\omega_1n_1+\hbar\omega_2n_2
\end{pmatrix}\nonumber \\
&=&\hbar\left[\left(\omega_1+\omega_2\right)j+\left(\omega_1-\omega_2\right)m_z
\right]I_{2\times 2}
+\frac{\hbar}{2}\begin{pmatrix}
-\delta & \Omega_R\sqrt{(2j+1)^2-4m_z^2} \cr
\Omega_R\sqrt{(2j+1)^2-4m_z^2} & \delta\cr
\end{pmatrix},
\label{2x2block}
\end{eqnarray}
\end{widetext}
where the two-photon detuning is defined to be 
$\delta\equiv\left(\omega_1-\omega_2\right)-\tilde{\omega}_0
\approx\Delta_1-\Delta_2$ and the number of
photons in each mode is written in terms of the pseudospin quantum numbers as
$n_1=j+m_z+\frac{1}{2}$, and $n_2=j-m_z-\frac{1}{2}$. Defining 
\begin{equation}
\Delta_{j,m_z}\equiv\sqrt{\Omega_R^2\left[(2j+1)^2-4m_z^2\right]+\delta^2},
\end{equation}
the eigenvalues of the two-dimensional blocks~(\ref{2x2block}) are 
\begin{equation}
E^{\pm}_{j,m_z}=\hbar[(\omega_1+\omega_2)j+(\omega_1-\omega_2)m_z]
\pm\frac{\hbar\Delta_{j,m_z}}{2}.
\label{Epm}
\end{equation}
Defining the mixing angle 
$\theta_{j,m_z}\equiv\cos^{-1}(-\delta/\Delta_{j,m_z})$, the 
dressed states (or polariton states) can be written 
\begin{equation}
\begin{pmatrix}
\left|u^+_{q_z,j,m_z}\right\rangle \cr
\noalign{\vskip0.1cm}
\left|u^-_{q_z,j,m_z}\right\rangle \cr
\end{pmatrix}
=\begin{pmatrix}
\cos\frac{\theta_{j,m_z}}{2} & \sin\frac{\theta_{j,m_z}}{2}\cr
\noalign{\vskip0.1cm}
-\sin\frac{\theta_{j,m_z}}{2} & \cos\frac{\theta_{j,m_z}}{2}\cr
\end{pmatrix}
\begin{pmatrix}
|q_z,\uparrow,j,m_z\rangle\cr
\noalign{\vskip0.1cm}
|q_z,\downarrow,j,m_z\rangle\cr
\end{pmatrix}.
\label{eq:mixing}
\end{equation}

Note that this treats 
$\tilde{\omega}_0$ and therefore $\delta$ as a constant independent of $j$ and
$m_z$, which is not strictly correct. Using results found in the Appendix, the
Stark-shifted atomic transition frequency is
\begin{eqnarray}
\tilde{\omega}_0&=&\omega_0+\frac{2g_{be}^2}{\Delta_2}\left(j-m_z\right)
-\frac{2g_{ae}^2}{\Delta_1}\left(j+m_z+1\right)
\nonumber \\
&=&\omega_0-\frac{2g_{ae}^2}{\Delta_1}+2j\left(\frac{g_{be}^2}{\Delta_2}
-\frac{g_{ae}^2}{\Delta_1}\right)\nonumber \\
&-&2m_z\left(\frac{g_{ae}^2}{\Delta_1}+\frac{g_{be}^2}{\Delta_2}\right)
\label{omegatwiddle}
\end{eqnarray}
where $\omega_0=(E_b-E_a)/\hbar$. The terms proportional to $j$ and $m_z$ can
be ignored to a first approximation. While $2j=n$ is a large number when many
photons occupy both modes, for a judicious choice of levels one should be able
to ensure that $g_{be}^2/\Delta_2\approx g_{ae}^2/\Delta_1$. Likewise, 
$m_z=(n_1-n_2-1)/2$, which should be small if $n_1\sim n_2$. Thus
$\tilde{\omega}_0\approx\omega_0-2g_{ae}^2/\Delta_1$ for each block. In fact,
as shown below, synthetic magnetic fields are maximized when $n_1\sim n_2$.
Even if $m_z$ is not small, for sufficiently big frequency detunings $\Delta_i$,
the second and the last term in Eq.~(\ref{omegatwiddle}) will be negligible and 
$\tilde{\omega}_0\approx\omega_0$. Yet the important off-diagonal term in the 
$2\times 2$ Hamiltonian blocks~(\ref{2x2block}) will remain appreciable as long 
as $j\gg m_z$. 

The generalized Jaynes-Cummings Hamiltonian $H_{\rm GJC}$ is now 
diagonal in the dressed state basis
\begin{equation} 
H_{\rm GJC}=\sum_{j,m_z,\tau}E_{j,m_z}^{\tau}P_{j,m_z,\tau}^\dag
P_{j,m_z,\tau},
\label{eq:HJC-polariton-basis}
\end{equation}
where $j=0,\frac{1}{2},1,\frac{3}{2},\ldots$, 
$m_z=-j-\frac{1}{2},\ldots,j+\frac{1}{2}$ in integer steps, and 
$\tau=\{0,\pm\}$. The energies $E^{\tau}_{j,m_z}$ are
defined in Eqs.~(\ref{E0}) and (\ref{Epm}). The polariton field creation 
operator is defined as (c.f.\ Ref.~\cite{Angelakis-2007,Koch-2009})
\begin{equation} 
P_{j,m_z,\tau}^\dagger=\ket{u_{q_z,j,m_z}^{\tau}}\bra{u_{q_z,0,-1/2}^0},
\end{equation}
where the dependence of the field operator on $q_z$ is suppressed for 
convenience.


\subsection{Diagonalizing the full Hamiltonian}
\label{subsec:FullH}

We are now in a position to obtain the matrix elements of the full 
Hamiltonian~(\ref{eq:Hamiltonian-Using-j-s}) in the dressed-state basis. It
suffices to obtain the coefficients 
\begin{equation}
t^{\tau\tau'}_{j,m_z}= k \bra{u^{\tau}_{q_z,j,m_z}} \sigma_z \ket{u^{\tau'}_{q_z,j,m_z}},
\label{eq:kinetic-term-effect-polariton-basis}
\end{equation}
which are
\begin{eqnarray} 
&&t_{j,m_z}^{\pm\pm}=\pm k\cos\theta_{j,m_z};\nonumber \\
&&t_{j,m_z}^{\pm\mp}=-k\sin\theta_{j,m_z};\nonumber \\
&&t^{00}_{j,j+\frac{1}{2}}=-t^{00}_{j,-j-\frac{1}{2}}=k.
\label{eq:coefficients}
\end{eqnarray}
With Eqs.~(\ref{E0}), (\ref{Epm}), (\ref{eq:HJC-polariton-basis}) and 
(\ref{eq:kinetic-term-effect-polariton-basis}) and some straightforward
algebra, the total Hamiltonian~(\ref{eq:Hamiltonian-Using-j-s}) becomes: 
\begin{eqnarray}
\tilde{H}_{\rm eff}&=&\sum_{j,m_z}\Bigg[\frac{\hbar^2}{2m}\Big(\tilde{q}_z + \sum_{\tau,\tau'}
t^{\tau\tau'}_{j,m_z}P^{\dag}_{j,m_z,\tau}P_{j,m_z,\tau'}\Big)^2
\nonumber \\
&+&\sum_{\tau}E^{\tau}_{j,m_z}P^{\dag}_{j,m_z,\tau}P_{j,m_z,\tau}\Bigg].
\label{eq:Hamiltonian-polariton-basis}
\end{eqnarray}
in the polariton basis. Here we have introduced the Doppler-shifted 
center-of-mass momentum $\tilde{q}_z \equiv  q_z-\Delta k/2$. 
The second term in the square brackets can be 
considered as a Zeeman energy shift for each sub-manifold. The first term 
contains the essential feature of the spin-orbit interaction: a spin-dependent
shift of the center-of-mass momentum. This can be made more explicit by 
introducing the effective spin operators
\begin{eqnarray}
X_{j,m_z}&\equiv&P_{j,m_z,+}^{\dag}P_{j,m_z,-}+P^{\dag}_{j,m_z,-}P_{j,m_z,+};
\nonumber \\
Z_{j,m_z}&\equiv&P^{\dag}_{j,m_z,+}P_{j,m_z,+}-P^{\dag}_{j,m_z,-}P_{j,m_z,-},
\end{eqnarray}
whenever $m_z\neq\pm(j+1/2)$. The Hamiltonian~(\ref{eq:Hamiltonian-polariton-basis}) 
can then be written
\begin{eqnarray}
\tilde{H}_{\rm eff}^{j,m_z}&=&\frac{\hbar^2}{2m}\Big[ \tilde{q}_z I_{j,m_z}+k\cos\theta_{j,m_z}Z_{j,m_z}\nonumber \\
&-&k\sin\theta_{j,m_z}X_{j,m_z}\Big]^2+\frac{\hbar\Delta_{j,m_z}}{2}Z_{j,m_z}\nonumber \\
&+&\hbar[(\omega_1+\omega_2)j+(\omega_1-\omega_2)m_z]I_{j,m_z},
\label{eq:Hamiltonian-polariton-basis2}
\end{eqnarray}
for some arbitrary $j\neq0$ and $m_z\neq\pm(j+1/2)$. 
This equation can be considered to be the main result of the present work. The
term in brackets corresponds to the Hamiltonian of a particle with a
Doppler-shifted center-of-mass momentum $\tilde{q}_z$
subject to a spin-dependent gauge field (i.e.\ the sign and strength of the 
guage field depends on the eigenstate of $m_s$). This takes a particularly 
simple form when $\cos\theta_{j,m_z}=0$, or zero two-photon
detuning $\delta=0$. 
For this case, the kinetic energy term takes the usual spin-orbit form 
$(\hbar^2/2m)[\tilde{q}_z I_{j,m_z}-kX_{j,m_z}]^2$. The last term is an 
overall energy shift for each two-dimensional block labeled by $m_z$. The 
penultimate term can be considered as a Zeeman splitting for 
the dressed states.

Eq.~(\ref{eq:Hamiltonian-polariton-basis2}) can be simplified slightly by
defining $E_R\equiv\hbar^2k^2/2m$ as the atomic recoil energy and
$p_z\equiv\tilde{q}_z/k$ as the Doppler-shifted center-of-mass momentum in 
units of $k$.  Ignoring the constant shift for each submanifold 
labeled by $m_z$, one obtains 
\begin{eqnarray}
\frac{\tilde{H}_{\rm eff}^{j,m_z}}{E_R}&=&
\Big[p_z I_{j,m_z}+\cos\theta_{j,m_z}Z_{j,m_z}
-\sin\theta_{j,m_z}X_{j,m_z}\Big]^2
\nonumber \\
&+&\frac{\hbar\Delta_{j,m_z}}{2E_R}Z_{j,m_z}.
\label{eq:H-in-Pauli-Basis}
\end{eqnarray}
This Hamiltonian can then be diagonalized, yielding the
dispersion relation
\begin{equation}
\frac{\epsilon^{\pm}_{j,m_z}(p_z)}{E_R}=p_z^2+1\pm\sqrt{4p_z^2
-2p_z\frac{\hbar\delta}{E_R}+\left(\frac{\hbar\Delta_{j,m_z}}{2E_R}
\right)^2}.
\label{eq:dispersion}
\end{equation}
Note that the energy dispersions for the 1D submanifolds $m_z=\pm(j+1/2)$
are independent of $j$ and $m_z$ and given by
\begin{align}
\frac{\epsilon^0_{j,\pm(j+1/2)}(p_z)}{E_R}=(p_z\pm1)^2,
\label{eq:dispersion-singlets}
\end{align}
where here also the energy offsets, Eq.~\eqref{E0}, have been omitted.

%
%

\section{Synthetic spin-orbit interactions and external magnetic fields}
\label{sec:syntheticfields}

\subsection{Synthetic spin-orbit interactions}
\label{subsec:SO}

To see the effect of the spin-orbit interactions, consider first the lower 
energy band in the simplest case of zero two-photon
detuning, $\delta=0$. This corresponds
to dressed energy levels that are equal superpositions of the original atomic
pseudospin states, c.f.\ Eq.~(\ref{eq:mixing}). The extrema of the dispersion 
relation $\partial\epsilon^-_{j,m_z}/\partial p_z=0$ are located at 
\begin{equation}
p_z^{\rm ex}=\left\{0,\pm\sqrt{1-\frac{1}{16}\left[(2j+1)^2-4m_z^2\right]\left(\frac{\hbar\Omega_R}{E_R}\right)^2}
\right\}.
\end{equation}
The non-zero solutions will be real only if 
\begin{equation} 
\frac{\hbar\Omega_R}{E_R}\leq\frac{4}{\sqrt{(2j+1)^2-4m_z^2}}.
\label{eq:non-zero-solutions}
\end{equation}
The largest possible value corresponds to $m_z^2=\left(j-\frac{1}{2}\right)^2$, 
which yields $\hbar\Omega_R\leq \sqrt{2/j}E_R$. The maximum number of photons 
in the 
cavity is therefore $n_{\rm max}=2j_{\rm max}=\lfloor 4(E_R/\hbar\Omega_R)^2
\rfloor$. The value of $n_{\rm max}$ can be made
arbitrarily large by setting $\hbar\Omega_R/E_R\to 0$, which corresponds to 
big frequency detunings $\Delta_i$ of the cavity mode frequencies from the 
atomic transitions (note that one cannot strictly set $\Omega_R=0$ unless the 
number of photons is exactly zero). The curvature at the extremum 
$p_z^{\rm ex}=0$ is given by 
\begin{equation}
\left.\frac{\partial^2\epsilon^-_{j,m_z}}{\partial p_z^2}
\right|_{p_z=0}=2-\frac{8}{\sqrt{(2j+1)^2-4m_z^2}}
\left(\frac{E_R}{\hbar\Omega_R}\right),
\end{equation}
which is negative for all $j<j_{\rm max}$; likewise, the curvature at the 
other two extrema is strictly positive. 

\begin{figure}[t]
\centering
\includegraphics [width=0.49\textwidth]{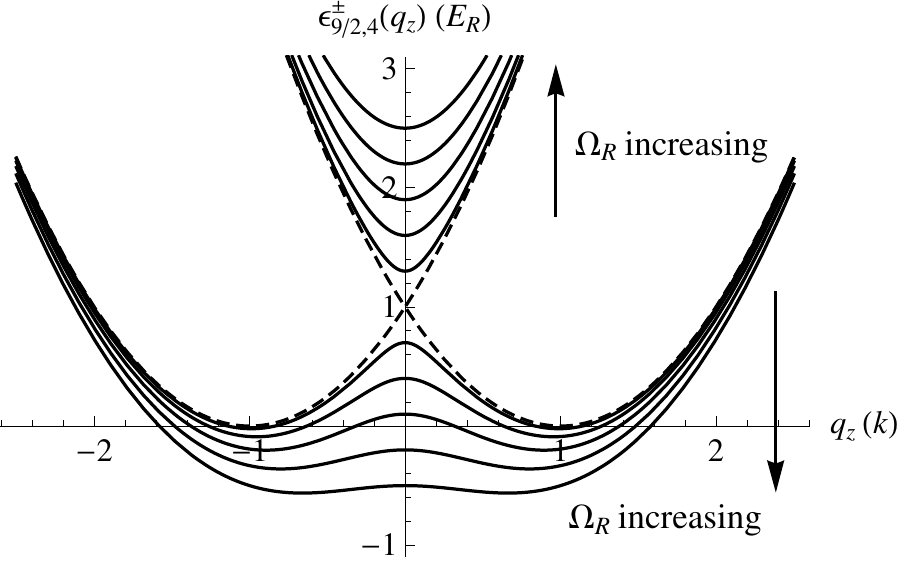}
\caption{The energy dispersion $\epsilon_{9/2,4}^{\pm}(q_z)$ is shown for 
$\delta=0$ and $\hbar\Omega_R/E_R$ in a range from $0$ to $0.5$ in equal 
increments. Increasing $\hbar\Omega_R/E_R$ (with the dashed curve being 
correspond to $\Omega_R=0$) reduces the barrier between the two minima in the 
energy dispersion.}
	\label{fig:Varying-Omega-R}
\end{figure}

The low-lying excitations for the resonant case, given by the band 
$\epsilon_{j,m_z}^-(p_z)$ in Eq.~(\ref{eq:dispersion}) with $\delta=0$,
therefore consist of a symmetric double-well centered at 
$\tilde{q}_z=q_z-\Delta k/2$ whose minima are located at $\tilde{q}_z\pm k$ in the limit 
$\hbar\Omega_R/E_R\to 0$. In this same limit the energy barrier reaches its 
maximum value of $E_R$. The exact energy bands $\epsilon^{\pm}_{j,m_z}(q_z)$ 
are depicted in 
Fig.~\ref{fig:Varying-Omega-R} for the particular case $\delta=0$, $j=9/2$, 
and $m_z=4$. For concreteness, we have used values for atomic mass and cavity wavenumbers 
corresponding to an $^{87}$Rb atom
confined in a ring cavity with nearly degenerate wavelength (i.e.\ 
$\Delta k\simeq0$) $\lambda=2\pi/k=804.1$~nm~\cite{Lin-2011-a}.
For large $\hbar\Omega_R/E_R$, the bottom of the dispersion curve 
is almost flat, but as $\hbar\Omega_R/E_R\to 0$ the minima approach a 
separation of $2k$ and are separated by a barrier approaching $E_R$. The 
existence of such a double well in the energy dispersion is a hallmark of
spin-orbit interactions, with the Hamiltonian minimized by two different 
dressed states $\ket{u^{\pm}_{q_z,j,m_z}}$.

\begin{figure}[t]
\centering
\includegraphics [width=0.49\textwidth]{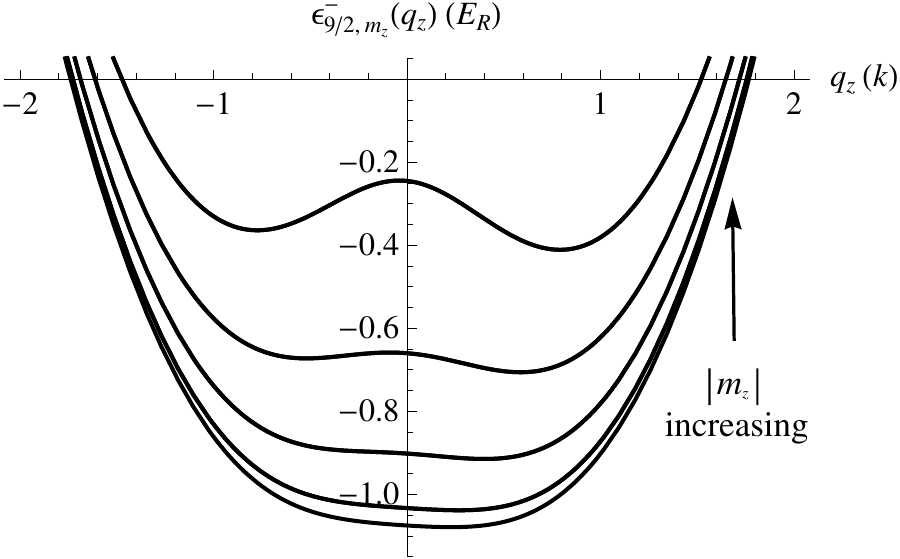}
\caption{The low-lying energy dispersions $\epsilon_{j,m_z}^-(q_z)$ are shown 
for $j=9/2$ and $|m_z|=0,...,4$.
Parameters correspond to $\hbar\Omega_R/E_R=0.415$ and $\hbar\delta/E_R=-0.06$.
The bottommost curve corresponds to $m_z=0$ and the topmost one to $m_z=\pm4$. 
For this choice, only the three topmost energy dispersions correspond to a 
spin-orbit interaction, with an appreciable energy barrier between minima only 
for $m_z=\pm4$.}
\label{fig:energy-dispersion-j-9/2-M-4}
\end{figure}

In the weakly non-resonant case $\delta\neq 0$ but $\hbar\delta/E_R\ll 1$, the 
double-well dispersion curves become asymmetric. For $\hbar\Omega_R/E_R\ll 1$,
the splitting $\gamma$ of the energy minima is approximately 
$$\gamma\approx\left[1-\frac{(2j+1)^2-4m_z^2}{32}
\left(\frac{\hbar\Omega_R}{E_R}\right)^2\right]\hbar\delta$$
which is independent of $j$ and $m_z$ only for $\Omega_R\to 0$. The minima are
now separated by $2k(\gamma/\hbar\delta)$.
Fig.~\ref{fig:energy-dispersion-j-9/2-M-4} depicts the atomic 
dispersion relations 
$\epsilon_{j,m_z}^-(q_z)$ for $j=9/2$ and $|m_z|=0,\ldots,4$, assuming 
$\hbar\delta/E_R=-0.06$ rather than zero, and 
$\hbar\Omega_R/E_R=0.415$. The bottommost curve corresponds to $m_z=0$ and the 
topmost one to $m_z=\pm4$.
For these parameters with 
$\hbar\delta/E_R$ small, the dispersion curves qualitatively follow the
$\delta=0$ results above. The uppermost curves with $|m_z|=2,3,4$
now correspond to asymmetric double-wells centered near $\tilde{q}_z=0$ with well 
minima slightly less than $2k$ apart and an energy splitting of order 
$\hbar\delta$ that is only weakly $m_z$-dependent. Only for the energy 
dispersion corresponding to $|m_z|=4$ is the energy barrier appreciable between 
the two minima. A single well results 
for $|m_z|=0,1$ because the energy difference between the two minima exceeds 
the barrier height. The analog of Eq.~(\ref{eq:non-zero-solutions}) for the 
$\delta\neq 0$ case is
\begin{equation}
\frac{\hbar\Omega_R}{E_R}\leq\frac{4[(2j+1)^2-4m_z^2]}
{[(2j+1)^2-4m_z^2+\delta^2/\Omega_R^2]^{3/2}},
\end{equation}
which is equivalent to $m_z\Delta_{j,m_z}\leq 2\hbar k^2\sin^2\theta_{j,m_z}$.
Violating this condition results in a single well. Thus, the spin-orbit 
interaction persists for most values of $m_z$, but is strongest when there is 
a large difference between the number of photons in the two cavity modes.


\subsection{Synthetic magnetic fields}
\label{subsec:B}

\begin{figure}[t]
\centering
\includegraphics [width=0.49\textwidth]{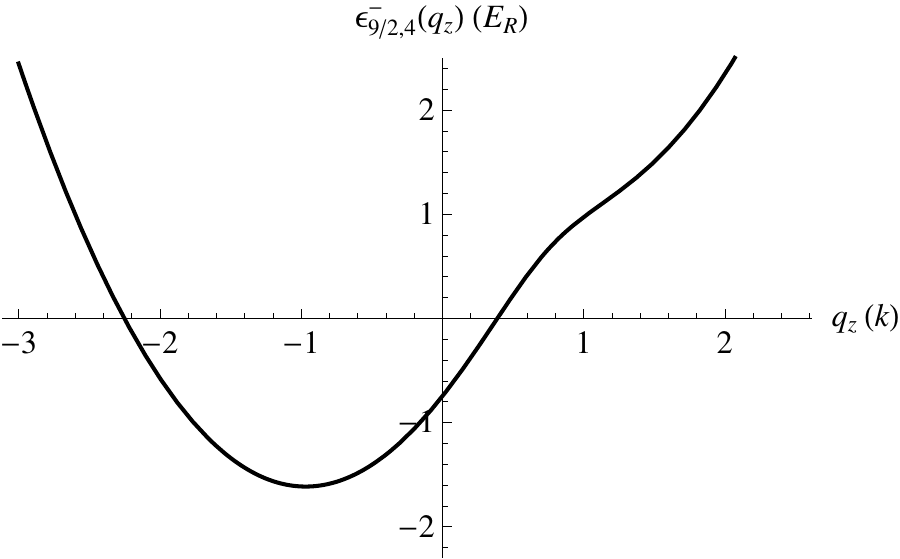}
\caption{The energy dispersion $\epsilon_{9/2,4}^{-}(q_z)$ is shown for 
$\hbar\Omega_R=0.3E_R$ and $\hbar\delta=3E_R$.}
\label{fig:Energy-Expension-Bad-Region}
\end{figure}

In the strongly non-resonant limit $\hbar\delta/E_R\gg 1$, there is only one
minimum of the dispersion curve $\epsilon_{j,m_z}^-$, located at 
\begin{equation} \label{strongly non-resonant minimum}
p_z^{\rm ex}\approx -1+\frac{(2j+1)^2-4m_z^2}{2}
\frac{\Omega_R^2}{\delta^2}.
\end{equation}
The lowest energy dispersion then consists of a single well, as shown in
Fig.~\ref{fig:Energy-Expension-Bad-Region}. For the parameters chosen ($j=9/2$, 
$m_z=4$, $k_1=k_2=k$, $\hbar \Omega_R=0.3 E_R$, and $\hbar\delta=3E_R$), the 
theoretical minimum of the dispersion curve based on the expression above 
occurs at $q_z=-0.82k$, which is close to the exact result $-0.97k$. These
parameters yield a mixing angle $\theta_{\frac{9}{2},4}\approx 0.21\pi$, 
indicating that the spin mixing is nevertheless appreciable. Note that the 
$\hbar\delta/E_R\gg 1$ condition is already well-satisfied here for the case
$\hbar\delta/E_R=3$.

Under these circumstances it is reasonable to also assume that 
$\delta\gg\Omega_R$ so that $\Delta_{j,m_z}\sim\delta$. The effective 
Hamiltonian~(\ref{eq:H-in-Pauli-Basis}) then becomes
\begin{equation}
\frac{\tilde{H}^{j,m_z}_{\rm eff}}{E_R}\approx\Big[p_z
I_{j,m_z}-Z_{j,m_z}\Big]^2+\frac{\hbar\delta}{2E_R}Z_{j,m_z}.
\end{equation}
The lower branch has dispersion relation 
\begin{equation}
\epsilon^{-}_{j,m_z}(p_z)\approx E_R\left[(p_z+1)^2
-\frac{\hbar\delta}{2E_R}\right],
\end{equation}
consistent with Eq.~\eqref{strongly non-resonant minimum} in the limit
$\delta\gg\Omega_R$. In terms of the original atomic momentum 
the dispersion relation becomes
\begin{eqnarray}
\epsilon^-_{j,m_z}&\approx&\frac{\hbar^2}{2m}\left(q_z-\frac{\Delta k}{2}+k\right)^2
-\frac{\hbar\delta}{2}\nonumber \\
&=&\frac{\hbar^2}{2m}\left(q_z+k_2\right)^2
-\frac{\hbar\delta}{2}.
\end{eqnarray}
Ignoring the overall energy shift $-\hbar\delta/2$, the dispersion relation is
equivalent to the canonical minimal coupling energy 
$\hbar^2(q_z-e^*A^*_z/\hbar)^2/2m$ of a particle with effective charge $e^*$ 
subject to a synthetic magnetic gauge potential $e^*A^*_z/\hbar
=-k_2$. This is simply $-k$ in the case $k_1=k_2$. 

Note that in the strongly non-resonant limit for negative two-photon detuning, 
that is $\hbar \delta/E_R\ll-1$, the minimum of the energy dispersion 
$\epsilon_{j,m_z}^-$ is instead located at
\begin{equation}
p_z^{\rm ex}\approx 1-\frac{(2j+1)^2-4m_z^2}{2}\frac{\Omega_R^2}{\delta^2}.
\end{equation}
The synthetic gauge potential then becomes $e^*A^*_z/\hbar=\Delta k/2+k
=k_1$, which can be considered as the artificial magnetic field for 
the other pseudospin dressed state. Thus the difference in the effective 
magnetic field strengths for the two pseudospin states is set by the maximum 
two-photon momentum transfer $k_1+k_2$, consistent with the continuum 
case~\cite{Lin-2009}.


The synthetic gauge potential 
$e^*A^*_z$ is position-independent and 
therefore cannot yield a synthetic magnetic field. Unfortunately it is not 
possible to make 
$k$ or $\Delta k$ depend on position. Instead, one can relax
the assumption that $\Omega_R/\delta\approx 0$ and rather make the two-photon
detuning $\delta$ position-dependent by applying a real external magnetic 
field gradient transverse to the cavity mode direction. For example, huge 
magnetic field gradients $B'$ are generated by integrating copper wires in the 
immediate vicinity of high-finesse optical cavities on microfabricated atom 
chips~\cite{Brahms-2011}. 

Consider a magnetic field gradient aligned along $\hat{y}$ giving rise to a 
position-dependent cavity detuning $\delta-\mu B'y/\hbar$, 
where $\mu/\hbar$ is the atomic gyromagnetic ratio. Taking the curl of 
Eq.~\eqref{strongly non-resonant minimum} then yields the synthetic magnetic 
field 
\begin{eqnarray}
\frac{e^*{\bf B}^*}{\hbar k}&=&\frac{\mu B'}{\hbar}\left[(2j+1)^2-4m_z^2\right]
\frac{\Omega_R^2}{(\delta-\mu B'y/\hbar)^3}\hat{x}\nonumber \\
&=&4\frac{\mu B'}{\hbar}n_1(n_2+1)\frac{\Omega_R^2}{(\delta-\mu B'y/\hbar)^3}
\hat{x},
\label{eq:artificial-magnetic-field}
\end{eqnarray}
which the second line is written in terms of the cavity occupation numbers. This
result shows that the magnitude of the synthetic magnetic field depends not 
only on the strength of the external magnetic field gradient $B'$ but also on the
population of the cavity modes, with the maximum corresponding to $n_1=(n+1)/2$
where $n$ is the total number of photons in the cavity. The maximum synthetic 
magnetic field therefore scales as $n^2$, which implies that much higher
synthetic magnetic fields for atoms can be reached in cavities than in the 
continuum. For example, choosing the same parameters as in 
Fig~\ref{fig:Energy-Expension-Bad-Region}, namely $j=9/2$, $m_z=4$ (or 
$n=9$ photons in the cavity with $n_1=9$, $n_2=0$, and spin down), 
$\hbar\delta/E_R=3$, $\hbar\Omega_R/E_R=0.3$, $\lambda=804.1$~nm~\cite{Lin-2009}
and 
$\mu B'/h=114$~kHz/$\mu$m~\cite{Brahms-2011}, gives a synthetic magnetic field 
of $\left|e^*B^*_x\right|\approx 3.8\hbar k/\mu$m at the cavity center. Instead
using the optimal value $j=9/2$, $m_z=0$ (or $n_1=5$, $n_2=4$, and spin down) 
gives $\left|e^*B^*_x\right|\approx 10\hbar k/\mu$m.

To get a sense of the magnitude of the synthetic magnetic 
field~(\ref{eq:artificial-magnetic-field}), consider the phase 
$\varphi$ acquired by the atomic wavefunction for a closed trajectory in the 
$yz$-plane. For concreteness, suppose that the path is a rectangle with lengths
$y_0$ and $z_0$. The accumulated phase is then
\begin{eqnarray}
\varphi&=&\oint\frac{e^*}{\hbar}{\bf A}^*\cdot d{\bf r}\nonumber \\
&=&2k z_0 n_1(n_2+1)\left(\frac{\Omega_R}{\delta}\right)^2\left[
\frac{1}{(1-\mu B'y_0/\hbar\delta)^2}-1\right].\nonumber \\
\end{eqnarray}
A natural choice is $y_0,z_0=\lambda/2$, corresponding to the length of one
unit cell of an external optical lattice generated by external lasers with
wavelength $\lambda$. Using the parameters above that maximize the synthetic 
magnetic field, one obtains $\varphi\approx 0.45\pi$. This is equivalent to
almost one quarter of a flux quantum per plaquette. Increasing the number of 
photons in the cavity to only $n=15$ increases the effective field to over one
flux quantum per plaquette. Comparable magnetic field strengths are impossible 
to attain in traditional condensed matter systems, requiring applied fields on 
the order of $B\sim 10^9$~G~\cite{Fisher-1985} while the highest 
non-destructive value so far achieved is just over 
$10^6$~G~\cite{Sebastian-2012}.

It is also instructive to compare the magnetic 
field~\eqref{eq:artificial-magnetic-field} with its continuum counterpart 
$q^*_LB^*_{Lx}/\hbar k = \hbar\delta_L'/(4E_L-\hbar\Omega_L)$ for low-lying 
band~\cite{Spielman-2009}. Here $\delta_L'$ is the detuning gradient related 
to an applied external magnetic field gradient, $\Omega_L$ is the laser 
two-photon Rabi frequency and the subscript $L$ refers to the laser based 
scheme. 
In order to have a consistent comparison, assume that 
$\delta_L'\approx \mu B'/\hbar$. The ratio between the two magnetic fields at 
the origin is then
$\zeta_{j,m_z}=4n_1(n_2+1)(4E_L/\hbar-\Omega_L)\Omega_R^2/\delta^3$.
For $\hbar \delta= 10 \hbar \Omega_R=3 E_R$ and 
$\hbar\Omega_L=16E_L$~\cite{Spielman-2009} (note that we have set $E_R=E_L$ for
convenience), the absolute value of $\zeta_{j,m_z}$ scales as $0.16n_1(n_2+1)$.
With only $n_1=n_2=25$ photons in each cavity mode, the artificial magnetic 
field in the cavity exceeds that in the continuum by over two orders of
magnitude.


\subsection{Cavity coherent states}
\label{subsec:coherent-state}

The foregoing analysis has assumed that the cavity modes are prepared in number
(or Fock) states $\ket{n}$. There have been several theoretical proposals for
quantum optics schemes that can deterministically prepare such Fock states in 
cavities~\cite{Parkins-1993, Parkins-1995, Gogyan-2012}. In these schemes, the 
maximum value of $n$ is restricted by the number of Zeeman sublevels of the
atom. In principle, the ideas can also be extended to the two-mode ring-cavity
states on which the present calculations are based.

That said, in the majority of experiments the cavity modes are best described
by being occupied by photon coherent states
\begin{align}
\ket{\alpha_i}=e^{-|\alpha_i|^2/2}\sum_{n_i=0}^{\infty}\frac{\alpha_i^{n_i}}
{\sqrt{n_i!}}\ket{n_i}, \quad i=1,2,\ldots,
\end{align}
where $|\alpha_i|^2=\langle n_i\rangle$ is the average number of photons in the
$i$th cavity mode coherent state. The probability of finding the $i$th mode 
in a particular photon number state $\ket{n_i}$ is then found using a Poisson 
distribution~\cite{Scully-1997}:
\begin{align}
P(n_i)=\frac{\langle n_i\rangle^{n_i} e^{-\langle n_i\rangle}}{n_i!}.
\end{align}
The dispersion curves for coherent states can then be obtained by summing over
all the Fock-state low-lying bands~\eqref{eq:dispersion} and 
\eqref{eq:dispersion-singlets}, weighted by their respective probabilities:
\begin{align}
\overline{\epsilon}(p_z)=\sum_{j,m_z}\sum_{\tau=\{0,-\}} P_{j,m_z}^{\tau} \epsilon_{j,m_z}^{\tau}(p_z),
\label{eq:average-energy-dispersion}
\end{align}
with the associated probabilities given by 
\begin{align}
P_{j,-(j+1/2)}^0&=\frac{1}{2}e^{-\langle n_1\rangle}
\frac{\langle n_2\rangle^{2j}e^{-\langle n_2\rangle}}{(2j)!};\nonumber\\
P_{j,j+1/2}^0&=\frac{1}{2}e^{-\langle n_2\rangle}\frac{\langle n_1\rangle^{2j}
e^{-\langle n_1\rangle}}{(2j)!},
\label{eq:prob0}
\end{align}
for $m_z=\pm(j+1/2)$ and
\begin{align}
P_{j,m_z}^-=&\frac{1}{2}e^{-(\langle n_1\rangle+\langle n_2\rangle)}
\Bigg[\frac{\langle n_1\rangle^{j+m_z+1/2}\langle n_2\rangle^{j-m_z-1/2}}
{(j+m_z+\frac{1}{2})!(j-m_z-\frac{1}{2})!}\nonumber\\
+&\frac{\langle n_1\rangle^{j+m_z-1/2}\langle n_2\rangle^{j-m_z+1/2}}
{(j+m_z-\frac{1}{2})!(j-m_z+\frac{1}{2})!}\Bigg],
\label{eq:prob-}
\end{align}
for the remaining states.

\begin{figure}[t]
\centering
\includegraphics [width=0.49\textwidth]{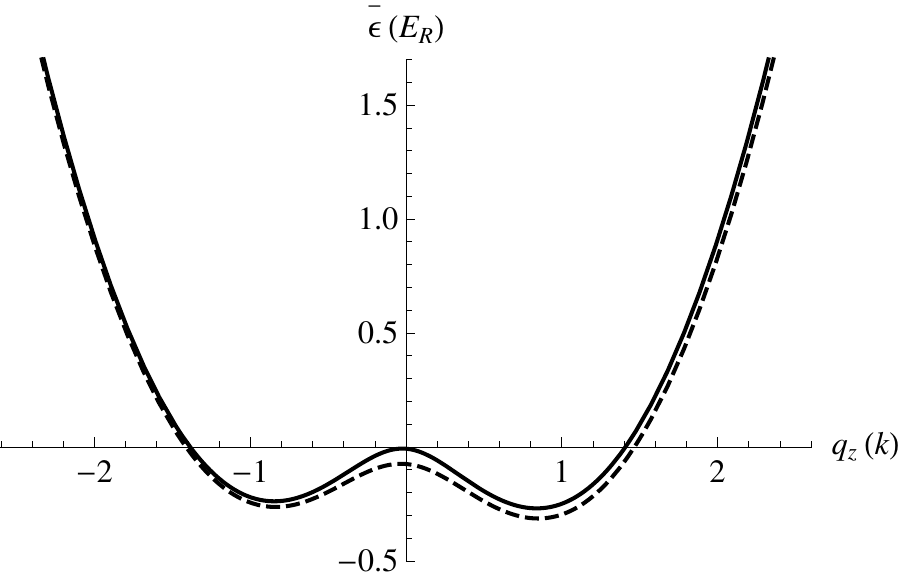}
\caption{The average energy dispersion $\overline\epsilon(p_z)$ is computed for 
$\langle n_1 \rangle=5$, $\langle n_2 \rangle=4$, 
$\hbar\Omega_R=0.215E_R$ and $\hbar\delta=-0.06E_R$. The dashed
curve corresponds to $\epsilon^-_{9/2,0}(p_z)$, see text.}
\label{fig:average-energy-soc}
\end{figure}

The coherent-state dispersion curve $\overline\epsilon(p_z)$, 
Eq.~\eqref{eq:average-energy-dispersion}, is shown as the solid curve in
Fig.~(\ref{fig:average-energy-soc}) for $\langle n_1 \rangle=5$, 
$\langle n_2 \rangle=4$, $\hbar\Omega_R=0.215E_R$ and $\hbar\delta=-0.06E_R$.
The single-manifold Fock-state energy dispersion $\epsilon^-_{j,m_z}$ with 
$j=(n_1+n_2)/2=9/2$ and $m_z=(n_1-n_2-1)/2=0$ (no need for averages with Fock 
states) is shown for comparison as the 
dashed curve. The results clearly show that the dispersion relations for 
coherent and Fock states are not appreciably different in the regime where 
the (exact or mean) occupations of the two cavity modes are comparable. Note 
that since $m_z=0$ corresponds to the shallowest spin-orbit interaction 
(see Fig.~(\ref{fig:energy-dispersion-j-9/2-M-4})), the value of $\Omega_R$ has 
been decreased to $0.215E_R$ in order to yield an appreciable barrier between 
the two minima. 

The coherent-state energy dispersion $\overline\epsilon(p_z)$ becomes 
increasingly distorted from that of a double-well as the average photon numbers 
in the two cavity modes become more asymmetric, i.e.\ for 
$\langle n_1 \rangle \gg \langle n_2 \rangle \sim0$ or vice versa, even in the 
case of zero two-photon detuning. This is because the probabilities
$P^0_{j,\pm(j+1/2)}$ and $P^0_{j,m_z}$ in Eqs.~(\ref{eq:prob0}) and 
(\ref{eq:prob-}) are proportional to $\langle n_i\rangle$. For 
$\langle n_1\rangle\to 0$, both $P^0_{j,m_z},P^0_{j,j+1/2}\to 0$ which favors
the occupation of the $m_z=-j-1/2$ singlet state. As discussed in 
Sec.~\ref{subsec:FullH}, the associated dispersion 
relation~\eqref{eq:dispersion-singlets} corresponds to a single well centered 
at $\tilde{q}_z=k$. For $\langle n_2\rangle\to 0$ the resulting dispersion 
relation for coherent states approaches a single well centered instead at 
$\tilde{q}_z=-k$. For $\langle n_1\rangle\lesssim\langle n_2\rangle$ or vice
versa, the double-well dispersion relation with $\delta=0$ can be made strongly 
asymmetric.  Thus, in the resonant and weakly non-resonant limits, more or less 
symmetric double-well dispersions can be realized for the coherent-state cavity 
modes provided that $\langle n_1 \rangle \sim \langle n_2 \rangle$ and 
$\hbar\Omega_R\ll E_R$.

\begin{figure}[t]
\centering
\includegraphics [width=0.49\textwidth]{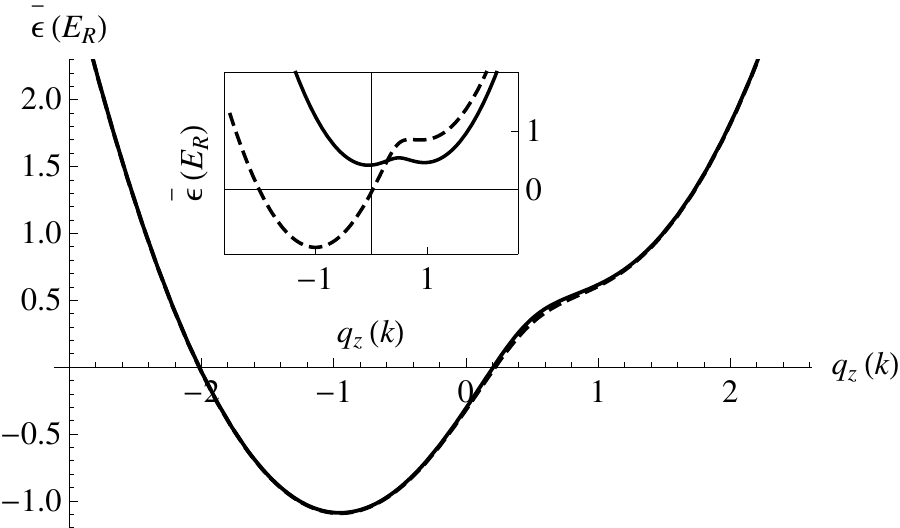}
\caption{The average energy dispersion $\overline\epsilon(p_z)$, computed for 
$\langle n_1 \rangle=8$, $\langle n_2 \rangle=7$, $\hbar\Omega_R=0.115E_R$ and $\hbar\delta=1.9E_R$. 
The dashed curve represents $\epsilon^-_{15/2,0}(p_z)$. 
Inset: $\langle n_1 \rangle=0.1$, $\langle n_2 \rangle=15$, with $\Omega_R$ and $\delta$ as the main panel.}
\label{fig:average-energy-restored}
\end{figure}

Just as was the case for spin-orbit interactions, for the analysis of synthetic 
magnetic fields for coherent-state cavity modes, one should again sum over all 
Fock states weighted by their probabilities, noting that that the singlet
manifolds $\epsilon^0_{j,\pm(j+1/2)}$ do not contribute in 
Eq.~\eqref{eq:artificial-magnetic-field}. Thus, the average ratio of the
cavity to continuum synthetic magnetic fields $\overline\zeta$ is now given by  
\begin{eqnarray}
\overline\zeta&=&4(\frac{\Omega_R^2}{\delta^3})\left(\frac{4E_L}{\hbar}-\Omega_L\right)\nonumber \\
&\times&\sum_{j,m_z} P_{j,m_z}^{-} 
\left(j+m_z+\frac{1}{2}\right)\left(j-m_z+\frac{1}{2}\right),
\end{eqnarray}
recalling that $n_1=j+m_z+\frac{1}{2}$ and $n_2=j-m_z-\frac{1}{2}$. When 
$\langle n_1 \rangle \simeq \langle n_2 \rangle$, the summation is
approximately equals to $\langle n_1 \rangle[\langle n_2 \rangle+1]$, and 
the the average ratio $\overline\zeta$ is approximately equals to the 
single-manifold ratio $\zeta_{j,m_z}$. This observation is borne out by
numerical calculations for the dispersion curve in the strongly non-resonant 
limit, as shown in Fig,~(\ref{fig:average-energy-restored}). The curves in the 
main panel correspond to the coherent-state (solid) and the single-manifold 
Fock-state (dashed) dispersions with $\langle n_1 \rangle=8$, 
$\langle n_2 \rangle=7$, $\hbar\Omega_R=0.115E_R$ and $\hbar\delta=1.9E_R$.
As evident from Fig,~(\ref{fig:average-energy-restored}), the coherent-state
and and single-manifold Fock-state dispersions are almost indistinguishable in 
this limit. 

It is interesting that in the strongly non-resonant limit one can restore 
symmetric single- or double-well dispersions by changing the average photon
numbers in the two cavity modes. This is illustrated in the inset of 
Fig,~(\ref{fig:average-energy-restored}), where $\langle n_1 \rangle=0.1$, 
$\langle n_2 \rangle=15$, with $\Omega_R$ and $\delta$ set to the same values
as the main panel (note that one cannot strictly set $\langle n_i \rangle$ to 
zero). The coherent-state energy dispersion (solid) is a shallow double well, 
with the centre of the double well displaced from the origin $q_z=0$ and the 
two minima located some fraction of $k$ apart from each other. For comparison,
the dashed curve represents the Fock-state energy dispersion with $n_1=0$,
$n_2=15$ and spin up (i.e. $j=15/2$ and $m_z=-7$). 
The change in shape has the same origins as the loss of the 
(approximately) symmetric double-well dispersion discussed above for the 
spin-orbit case: as $\langle n_1\rangle\to 0$, the occupations of all but the 
$m_z=-j-1/2$ singlet will be strongly suppressed, which will favor the
appearance of an additional well in the vicinity of $q_z=k$ and the suppression
of the mininum near $q_z=-k$. For very small values of $\langle n_i\rangle$,
the synthetic magnetic field (which has its origin strictly in the doublets)
will then approach zero.

%
%

\section{Discussion and Conclusions}
\label{sec:conclusions}

In this work, we considered three internal atomic states in the $\Lambda$ 
scheme coupled to two counter-propagating far off-resonance ring-cavity modes. 
After adiabatic elimination of the atomic excited state by virtue of the large 
detunings of the cavity frequencies from the atomic transitions, we obtained 
the effective Hamiltonian $\tilde{H}_{\rm eff}$. This Hamiltonian can be 
divided into two parts: a generalized Jaynes-Cummings Hamiltonian 
$H_{\rm GJC}$ and a kinetic contribution 
$\tilde{H}_{\rm eff}-H_{\rm GJC}$. Diagonalizing $H_{GJC}$ yields 
dressed (i.e.\ polariton) states, so the total Hamiltonian 
$\tilde{H}_{\rm eff}$ is then naturally expressed in the polariton basis, with 
$H_{\rm GJC}$ essentially a Zeeman shift for the polaritons. 

The dispersion relation of the total Hamiltonian $\tilde{H}_{\rm eff}$ is found 
to correspond to a symmetric double-well structure in the limit of zero 
two-photon detuning $\delta$,
which is the hallmark of an induced spin-orbit interaction. The energy barrier 
between degenerate polariton ground states is found to shrink as the Rabi 
frequency $\Omega_R$ increases. Furthermore, the strength of the spin-orbit 
interactions is enhanced by accentuating the asymmetry in the occupation of 
the two cavity modes. Assuming Fock states the largest energy barrier occurs 
for one photon in one mode, and all the other photons in the other mode. For
coherent states a strong asymmetry in the average mode occupations destroys
the double-well structure, instead yielding single-well dispersion relations,
so large barriers instead require approximately equal average photon numbers
in each cavity mode as well as smaller values of $\Omega_R$. In either case,
this mode occupation parameter is unique to cavities, with no analog in the 
continuum where atoms interact with many-photon laser fields, and is in 
practise an experimentally accessible parameter. For small two-photon 
detunings, the energy dispersions become slightly asymmetric; that is, one of 
the two energy wells is shifted up or down with respect to the other.

For larger cavity detunings a single well results, corresponding to a 
vector gauge potential for one pseudospin dressed state. In the presence of a 
real external magnetic field gradient, this potential becomes a synthetic 
magnetic field for the neutral atom. For large occupation asymmetry, the 
strength of the magnetic field is proportional to the number of photons in 
one of the modes, but the largest fields result for smallest asymmetry in which
case the strength is proportional to the square of the total number of cavity 
photons. For large magnetic field gradients, which can be generated particularly
easily with integrated atom-chip cavity QED, even moderate occupations (on the
order of 10-20 photons in the cavity) result in synthetic magnetic fields that
can easily exceed one flux quantum per cavity wavelength squared, much larger
than is accessible using (fundamentally weak-coupling) resonant Raman 
techniques in the continuum.

The present strong-coupling calculations have neglected cavity gain and loss 
that are non-negligible in many practical situations, such as the presence of 
cavity pump lasers and loss due to the spontaneous decay into vacuum modes or 
decay of the cavity modes. Under these conditions, the exact polariton approach
that is adopted here is not wholly suitable, and other approaches such as use 
of a master equation~\cite{Meystre1999} are required. That said, in the 
weak-coupling limit it should be possible to adiabatically eliminate the 
cavity fields to obtain an effective Hamiltonian for the atoms. This regime
will be explored in future work.

The natural emergence of spin-orbit interactions and strong synthetic magnetic 
fields for neutral atoms in ring cavities suggests that exotic quantum phases
would result for many atom systems. For example, one might expect topological 
insulators, including quantum Hall-type states, to result. Cavities provide a
unique environment where strong atom-atom correlations could emerge naturally.
The dynamic of the atomic field operators depends on the cavity fields and 
vice versa; in this respect, the system resembles a real material characterized 
by strong interplay between the electrons and phonons. In the presence of an 
additional optical lattice, for example, the effective Hamiltonian for the 
cavity atoms would resemble a spin-orbit-coupled Hubbard 
Hamiltonian~\cite{Kaplan-1983} locally, but would also enjoy a variety of 
infinite-range atom-atom interactions. These would include arbitrarily 
long-range density-density interaction of the form $n_{i,\sigma} n_{j,\sigma}$,
where $n_{i,\sigma}$ is the particle number operator for pseudospin state 
$\sigma\in\{a,b\}$ at the lattice site $i$. The emergence of such 
infinite-range interactions is a direct consequence of the back-action of the 
cavity fields in the atomic states and they drastically modify the quantum 
phases of the original Hubbard model and can give rise to exotic states 
many-particle \cite{Maschler-2008,Muller-2012}. These possibilities and 
related questions will be the focus of future work.

\begin{acknowledgments}
The authors are grateful to Barry Sanders and Paul Barclay for constructive 
criticisms. This work was supported by the Natural Sciences and Engineering 
Research Council of Canada.
\end{acknowledgments}
%
%
\appendix*
\section{Adiabatic Elimination}
\label{sec:Adiabatic-Elimination}

The procedure to adiabatically eliminate the atomic excited state in the
three-level Hamiltonian, Eq.~(\ref{eq:Hamiltonian-RWA}), is given in 
Ref.~\cite{Gerry-1990} and the derivation below follows this with some 
generalizations. The Heisenberg equations of motion for the atomic transition 
operators are given by 
\begin{eqnarray} 
i\hbar \dot{\sigma}_{ae}&=&E_{ea}\sigma_{ae}
+\hbar g_{ae}(z)a_1\sigma_z^{ae}+\hbar g_{be}(z)a_2\sigma_{ab};\nonumber\\
i\hbar \dot{\sigma}_{be}&=&E_{eb}\sigma_{be}+\hbar g_{be}(z)a_2\sigma_z^{be}
+\hbar g_{ae}(z)a_1\sigma_{ba};\qquad
\label{A-Heisenberg-Eq}
\end{eqnarray}
where $\sigma_z^{ij}\equiv\sigma_{ii}-\sigma_{jj}
=|i\rangle\langle i|-|j\rangle\langle j|$. Note that 
$\sigma_{ij}^{\dag}=\sigma_{ji}$ so equations of motion for these quantities 
follow directly from those above.
Defining new variables $\sigma_{ae}(t)\equiv\tilde{\sigma}_{ae}(t)
e^{-i\omega_1t}$, $\sigma_{be}(t)\equiv\tilde{\sigma}_{be}(t)e^{-i\omega_2t}$,
$a_1(t)\equiv\tilde{a}_1(t)e^{-i\omega_1t}$, 
$a_2(t)\equiv\tilde{a}_2(t)e^{-i\omega_2t}$, and 
$\sigma_{ab}(t)=\sigma_{ae}(t)\sigma_{eb}(t)\equiv\tilde{\sigma}_{ab}(t)
e^{i(\omega_2-\omega_1)t}$,
with $\tilde{\sigma}_{ab}=\tilde{\sigma}_{ae}\tilde{\sigma}_{eb}$, the Heisenberg equation of motions (\ref{A-Heisenberg-Eq}) take the form
\begin{eqnarray}
i\dot{\tilde{\sigma}}_{ae}&=&-\Delta_1\tilde{\sigma}_{ae}
+g_{be}(z)\tilde{a}_2\tilde{\sigma}_{ab}+g_{ae}(z)\tilde{a}_1\sigma_z^{ae};
\nonumber\\
i\dot{\tilde{\sigma}}_{be}&=&-\Delta_2\tilde{\sigma}_{be}
+g_{ae}(z)\tilde{a}_1\tilde{\sigma}_{ba}+g_{be}(z)\tilde{a}_2\sigma_z^{be};
\end{eqnarray}
where $\Delta_1=\omega_1-E_{ea}/\hbar$ and 
$\Delta_2=\omega_2-E_{eb}/\hbar$. The adiabatic condition 
$\hbar\Delta_j\gg E_{ba}$ implies that the time-dependence of all atomic 
transition operators involving the excited state is vanishingly small; that 
is, $\dot{\tilde{\sigma}}_{ae}=\dot{\tilde{\sigma}}_{ea}
=\dot{\tilde{\sigma}}_{be}=\dot{\tilde{\sigma}}_{eb}\approx 0$. This yields
\begin{eqnarray}
\tilde{\sigma}_{ae}&\approx&\frac{1}{\Delta_1}\left[g_{be}(z)\tilde{a}_2
\tilde{\sigma}_{ab}+g_{ae}(z)\tilde{a}_1\tilde{\sigma}_z^{ae}\right];
\nonumber\\
\tilde{\sigma}_{be}&\approx&\frac{1}{\Delta_2}\left[g_{ae}(z)\tilde{a}_1
\tilde{\sigma}_{ba}+g_{be}(z)\tilde{a}_2\tilde{\sigma}_z^{be}\right];
\label{eq:adiabatic-transition-operators}
\end{eqnarray} 
Because $\tilde{\sigma}_{ee}=\tilde{\sigma}_{ea}\tilde{\sigma}_{ae}
=\tilde{\sigma}_{eb}\tilde{\sigma}_{be}\propto|g|^2/\Delta^2\ll 1$ by 
assumption, all terms involving only the excited state can be neglected; thus
$\tilde{\sigma}_z^{ae}\approx\tilde{\sigma}_{aa}$ and
$\tilde{\sigma}_z^{be}\approx\tilde{\sigma}_{bb}$. The excited state of the 
atom is therefore decoupled from the other degrees of freedom in the 
Hamiltonian. Substituting Eq.~(\ref{eq:adiabatic-transition-operators}) into 
Eq.~(\ref{eq:Hamiltonian-RWA}) gives
\begin{eqnarray}
H_{\rm eff}&\approx&\frac{\hbar^2 q_z^2}{2m}I_{2\times 2}+\hbar(\omega_1\tilde{a}^\dagger_1
\tilde{a}_1+\omega_2\tilde{a}_2^\dagger\tilde{a}_2)\nonumber \\
&+&\Bigg[\hbar g_{ae}^*(z)g_{be}(z)\left(\frac{1}{\Delta_1}
+\frac{1}{\Delta_2}\right)\tilde{a}_1^{\dag}\tilde{a}_2\tilde{\sigma}_{ab}
+\mbox{H.c.}\Bigg]\nonumber \\
&+&\left[E_a+\frac{\hbar|g_{ae}|^2}{\Delta_1}\left(\tilde{a}_1
\tilde{a}_1^{\dag}+\tilde{a}_1^{\dag}\tilde{a}_1\right)\right]
\tilde{\sigma}_{aa}\nonumber \\
&+&\left[E_b+\frac{\hbar|g_{be}|^2}{\Delta_2}\left(\tilde{a}_2
\tilde{a}_2^{\dag}+\tilde{a}_2^{\dag}\tilde{a}_2\right)\right]
\tilde{\sigma}_{bb},
\end{eqnarray}
where the Hamiltonian for the excited state is completely ignored. Defining 
the AC Stark-shifted energies
\begin{eqnarray}
\tilde{E}_a&\equiv&E_a+\frac{2\hbar\left|g_{ae}\right|^2}{\Delta_1}\left(
\tilde{a}_1^{\dag}\tilde{a}_1+\frac{1}{2}\right);\nonumber\\
\tilde{E}_b&\equiv&E_b+\frac{2\hbar\left|g_{be}\right|^2}{\Delta_2}\left(
\tilde{a}_2^{\dag}\tilde{a}_2+\frac{1}{2}\right),\nonumber\\
\label{ACStark}
\end{eqnarray}
and the two-photon Rabi frequency 
\begin{equation}
\Omega_R\equiv g_{ae}g_{be}\frac{\Delta_1+\Delta_2}{\Delta_1\Delta_2},
\label{Rabi}
\end{equation}
where $\{g_{ae},g_{be}\}\in\Re$, the adiabatically-eliminated Hamiltonian is
\begin{eqnarray}
H_{\rm eff}&\approx&\frac{\hbar^2 q_z^2}{2m}I_{2\times 2}+\tilde{E}_a\tilde{\sigma}_{aa}
+\tilde{E}_b\tilde{\sigma}_{bb}+\hbar(\omega_1\tilde{a}^\dagger_1\tilde{a}_1
+\omega_2\tilde{a}_2^\dagger\tilde{a}_2)\nonumber \\
&+&\hbar\Omega_R\left(e^{i(k_1+k_2)z}\tilde{a}_2^{\dag}\tilde{a}_1
\tilde{\sigma}_{ba}+\mbox{H.c.}\right).\nonumber \\
\end{eqnarray}
Defining $\hbar\tilde{\omega}_0\equiv\tilde{E}_b-\tilde{E}_a$ and
$\hbar\overline{\omega}\equiv(\tilde{E}_a+\tilde{E}_b)/2$, then the Hamiltonian 
becomes
\begin{eqnarray}
H_{\rm eff}&\approx&\frac{\hbar^2 q_z^2}{2m}I_{2\times 2}+\frac{1}{2}\hbar\tilde{\omega}_0
\tilde{\sigma}_z^{ba}+\hbar(\omega_1\tilde{a}^\dagger_1\tilde{a}_1
+\omega_2\tilde{a}_2^\dagger\tilde{a}_2)\nonumber \\
&+&\hbar\Omega_R\left(e^{i(k_1+k_2)z}\tilde{a}_2^{\dag}\tilde{a}_1
\tilde{\sigma}_{ba}+\mbox{H.c.}\right),
\end{eqnarray}
where the constant term $\hbar\overline{\omega}\left(\tilde{\sigma}_{aa}
+\tilde{\sigma}_{bb}\right)=\hbar\overline{\omega}I_{2\times 2}$ has no 
effect on the dynamics and is therefore neglected. Because the 
frequency-dependent exponential factors all cancel, one can replace
$\tilde{\sigma}\to\sigma$ and $\tilde{a}\to a$ without loss of generality,
and this yields the effective Hamiltonian \eqref{eq:Hamiltonian-adiabatic}.

\bibliographystyle{apsrev}
\bibliography{sobib}

\begin{thebibliography}{49}
\expandafter\ifx\csname natexlab\endcsname\relax\def\natexlab#1{#1}\fi
\expandafter\ifx\csname bibnamefont\endcsname\relax
  \def\bibnamefont#1{#1}\fi
\expandafter\ifx\csname bibfnamefont\endcsname\relax
  \def\bibfnamefont#1{#1}\fi
\expandafter\ifx\csname citenamefont\endcsname\relax
  \def\citenamefont#1{#1}\fi
\expandafter\ifx\csname url\endcsname\relax
  \def\url#1{\texttt{#1}}\fi
\expandafter\ifx\csname urlprefix\endcsname\relax\def\urlprefix{URL }\fi
\providecommand{\bibinfo}[2]{#2}
\providecommand{\eprint}[2][]{\url{#2}}

\bibitem[{\citenamefont{Dresselhaus}(1955)}]{Dresselhaus-1955}
\bibinfo{author}{\bibfnamefont{G.}~\bibnamefont{Dresselhaus}},
  \bibinfo{journal}{Phys. Rev.} \textbf{\bibinfo{volume}{100}},
  \bibinfo{pages}{580} (\bibinfo{year}{1955}).

\bibitem[{\citenamefont{Bychkov and Rashba}(1984)}]{Rashba-1984}
\bibinfo{author}{\bibfnamefont{Y.~A.} \bibnamefont{Bychkov}} \bibnamefont{and}
  \bibinfo{author}{\bibfnamefont{E.~I.} \bibnamefont{Rashba}},
  \bibinfo{journal}{J. Phys. C} \textbf{\bibinfo{volume}{17}},
  \bibinfo{pages}{6039} (\bibinfo{year}{1984}).

\bibitem[{\citenamefont{Kane and Mele}(2005{\natexlab{a}})}]{Kane-2005-a}
\bibinfo{author}{\bibfnamefont{C.~L.} \bibnamefont{Kane}} \bibnamefont{and}
  \bibinfo{author}{\bibfnamefont{E.~J.} \bibnamefont{Mele}},
  \bibinfo{journal}{Phys. Rev. Lett.} \textbf{\bibinfo{volume}{95}},
  \bibinfo{pages}{226801} (\bibinfo{year}{2005}{\natexlab{a}}).

\bibitem[{\citenamefont{Kane and Mele}(2005{\natexlab{b}})}]{Kane-2005-b}
\bibinfo{author}{\bibfnamefont{C.~L.} \bibnamefont{Kane}} \bibnamefont{and}
  \bibinfo{author}{\bibfnamefont{E.~J.} \bibnamefont{Mele}},
  \bibinfo{journal}{Phys. Rev. Lett.} \textbf{\bibinfo{volume}{95}},
  \bibinfo{pages}{146802} (\bibinfo{year}{2005}{\natexlab{b}}).

\bibitem[{\citenamefont{Hasan and Kane}(2010)}]{Hasan-2010}
\bibinfo{author}{\bibfnamefont{M.~Z.} \bibnamefont{Hasan}} \bibnamefont{and}
  \bibinfo{author}{\bibfnamefont{C.~L.} \bibnamefont{Kane}},
  \bibinfo{journal}{Rev. Mod. Phys.} \textbf{\bibinfo{volume}{82}},
  \bibinfo{pages}{3045} (\bibinfo{year}{2010}).

\bibitem[{\citenamefont{Schnyder et~al.}(2008)\citenamefont{Schnyder, Ryu,
  Furusaki, and Ludwig}}]{Schnyder-2008}
\bibinfo{author}{\bibfnamefont{A.~P.} \bibnamefont{Schnyder}},
  \bibinfo{author}{\bibfnamefont{S.}~\bibnamefont{Ryu}},
  \bibinfo{author}{\bibfnamefont{A.}~\bibnamefont{Furusaki}}, \bibnamefont{and}
  \bibinfo{author}{\bibfnamefont{A.~W.~W.} \bibnamefont{Ludwig}},
  \bibinfo{journal}{Phys. Rev. B} \textbf{\bibinfo{volume}{78}},
  \bibinfo{pages}{195125} (\bibinfo{year}{2008}).

\bibitem[{\citenamefont{Wen}(2012)}]{Wen-2012}
\bibinfo{author}{\bibfnamefont{X.-G.} \bibnamefont{Wen}},
  \bibinfo{journal}{Phys. Rev. B} \textbf{\bibinfo{volume}{85}},
  \bibinfo{pages}{085103} (\bibinfo{year}{2012}).

\bibitem[{\citenamefont{Slager et~al.}(2013)\citenamefont{Slager, Mesaros,
  Juri\v{c}i\'{c}, and Zaanen}}]{Slager-2013}
\bibinfo{author}{\bibfnamefont{R.-J.} \bibnamefont{Slager}},
  \bibinfo{author}{\bibfnamefont{A.}~\bibnamefont{Mesaros}},
  \bibinfo{author}{\bibfnamefont{V.}~\bibnamefont{Juri\v{c}i\'{c}}},
  \bibnamefont{and} \bibinfo{author}{\bibfnamefont{J.}~\bibnamefont{Zaanen}},
  \bibinfo{journal}{Nature Phys.} \textbf{\bibinfo{volume}{9}},
  \bibinfo{pages}{98} (\bibinfo{year}{2013}).

\bibitem[{\citenamefont{Jaksch and Zoller}(2005)}]{Jaksch-2005}
\bibinfo{author}{\bibfnamefont{D.}~\bibnamefont{Jaksch}} \bibnamefont{and}
  \bibinfo{author}{\bibfnamefont{P.}~\bibnamefont{Zoller}},
  \bibinfo{journal}{Ann. Phys.} \textbf{\bibinfo{volume}{315}},
  \bibinfo{pages}{52} (\bibinfo{year}{2005}).

\bibitem[{\citenamefont{Lewenstein et~al.}(2007)\citenamefont{Lewenstein,
  Sanpera, Ahufinger, Damskic, Sen(De), and Sen}}]{Lewenstein-2007}
\bibinfo{author}{\bibfnamefont{M.}~\bibnamefont{Lewenstein}},
  \bibinfo{author}{\bibfnamefont{A.}~\bibnamefont{Sanpera}},
  \bibinfo{author}{\bibfnamefont{V.}~\bibnamefont{Ahufinger}},
  \bibinfo{author}{\bibfnamefont{B.}~\bibnamefont{Damskic}},
  \bibinfo{author}{\bibfnamefont{A.}~\bibnamefont{Sen(De)}}, \bibnamefont{and}
  \bibinfo{author}{\bibfnamefont{U.}~\bibnamefont{Sen}}, \bibinfo{journal}{Adv.
  Phys.} \textbf{\bibinfo{volume}{56}}, \bibinfo{pages}{243}
  (\bibinfo{year}{2007}).

\bibitem[{\citenamefont{Bloch et~al.}(2008)\citenamefont{Bloch, Dalibard, and
  Zwerger}}]{Bloch-2008}
\bibinfo{author}{\bibfnamefont{I.}~\bibnamefont{Bloch}},
  \bibinfo{author}{\bibfnamefont{J.}~\bibnamefont{Dalibard}}, \bibnamefont{and}
  \bibinfo{author}{\bibfnamefont{W.}~\bibnamefont{Zwerger}},
  \bibinfo{journal}{Rev. Mod. Phys.} \textbf{\bibinfo{volume}{80}},
  \bibinfo{pages}{885} (\bibinfo{year}{2008}).

\bibitem[{\citenamefont{Dalibard et~al.}(2011)\citenamefont{Dalibard, Gerbier,
  Juzeli\={u}nas, and \"{O}hberg}}]{Dalibard-2011}
\bibinfo{author}{\bibfnamefont{J.}~\bibnamefont{Dalibard}},
  \bibinfo{author}{\bibfnamefont{F.}~\bibnamefont{Gerbier}},
  \bibinfo{author}{\bibfnamefont{G.}~\bibnamefont{Juzeli\={u}nas}},
  \bibnamefont{and}
  \bibinfo{author}{\bibfnamefont{P.}~\bibnamefont{\"{O}hberg}},
  \bibinfo{journal}{Rev. Mod. Phys.} \textbf{\bibinfo{volume}{83}},
  \bibinfo{pages}{1523} (\bibinfo{year}{2011}).

\bibitem[{\citenamefont{Lin et~al.}(2009)\citenamefont{Lin, Compton,
  Jim\'{e}nez-García, Porto, and Spielman}}]{Lin-2009}
\bibinfo{author}{\bibfnamefont{Y.-J.} \bibnamefont{Lin}},
  \bibinfo{author}{\bibfnamefont{R.~L.} \bibnamefont{Compton}},
  \bibinfo{author}{\bibfnamefont{K.}~\bibnamefont{Jim\'{e}nez-García}},
  \bibinfo{author}{\bibfnamefont{J.~V.} \bibnamefont{Porto}}, \bibnamefont{and}
  \bibinfo{author}{\bibfnamefont{I.~B.} \bibnamefont{Spielman}},
  \bibinfo{journal}{Nature} \textbf{\bibinfo{volume}{462}},
  \bibinfo{pages}{628} (\bibinfo{year}{2009}).

\bibitem[{\citenamefont{Lin et~al.}(2011{\natexlab{a}})\citenamefont{Lin,
  Compton, Jim\'{e}nez-Garcia, Phillips, Porto, and Spielman}}]{Lin-2011-b}
\bibinfo{author}{\bibfnamefont{Y.-J.} \bibnamefont{Lin}},
  \bibinfo{author}{\bibfnamefont{R.~L.} \bibnamefont{Compton}},
  \bibinfo{author}{\bibfnamefont{K.}~\bibnamefont{Jim\'{e}nez-Garcia}},
  \bibinfo{author}{\bibfnamefont{W.~D.} \bibnamefont{Phillips}},
  \bibinfo{author}{\bibfnamefont{J.~V.} \bibnamefont{Porto}}, \bibnamefont{and}
  \bibinfo{author}{\bibfnamefont{I.~B.} \bibnamefont{Spielman}},
  \bibinfo{journal}{Nature Physics} \textbf{\bibinfo{volume}{7}},
  \bibinfo{pages}{531} (\bibinfo{year}{2011}{\natexlab{a}}).

\bibitem[{\citenamefont{Hatano et~al.}(2007)\citenamefont{Hatano, Shirasaki,
  and Nakamura}}]{Hatano-2007}
\bibinfo{author}{\bibfnamefont{N.}~\bibnamefont{Hatano}},
  \bibinfo{author}{\bibfnamefont{R.}~\bibnamefont{Shirasaki}},
  \bibnamefont{and} \bibinfo{author}{\bibfnamefont{H.}~\bibnamefont{Nakamura}},
  \bibinfo{journal}{Phys. Rev. A} \textbf{\bibinfo{volume}{75}},
  \bibinfo{pages}{032107} (\bibinfo{year}{2007}).

\bibitem[{\citenamefont{Lin et~al.}(2011{\natexlab{b}})\citenamefont{Lin,
  Jim\'{e}nez-Garcia, and Spielman}}]{Lin-2011-a}
\bibinfo{author}{\bibfnamefont{Y.-J.} \bibnamefont{Lin}},
  \bibinfo{author}{\bibfnamefont{K.}~\bibnamefont{Jim\'{e}nez-Garcia}},
  \bibnamefont{and} \bibinfo{author}{\bibfnamefont{I.~B.}
  \bibnamefont{Spielman}}, \bibinfo{journal}{Nature}
  \textbf{\bibinfo{volume}{471}}, \bibinfo{pages}{83}
  (\bibinfo{year}{2011}{\natexlab{b}}).

\bibitem[{\citenamefont{Moller and Cooper}(2012)}]{Cooper-2012-a}
\bibinfo{author}{\bibfnamefont{G.}~\bibnamefont{Moller}} \bibnamefont{and}
  \bibinfo{author}{\bibfnamefont{N.~R.} \bibnamefont{Cooper}},
  \bibinfo{journal}{Phys. Rev. Lett.} \textbf{\bibinfo{volume}{108}},
  \bibinfo{pages}{045306} (\bibinfo{year}{2012}).

\bibitem[{\citenamefont{Baur and Cooper}(2012)}]{Cooper-2012-b}
\bibinfo{author}{\bibfnamefont{S.~K.} \bibnamefont{Baur}} \bibnamefont{and}
  \bibinfo{author}{\bibfnamefont{N.~R.} \bibnamefont{Cooper}},
  \bibinfo{journal}{Phys. Rev. Lett.} \textbf{\bibinfo{volume}{109}},
  \bibinfo{pages}{265301} (\bibinfo{year}{2012}).

\bibitem[{\citenamefont{Cooper and Dalibard}(2013)}]{Cooper-2013}
\bibinfo{author}{\bibfnamefont{N.~R.} \bibnamefont{Cooper}} \bibnamefont{and}
  \bibinfo{author}{\bibfnamefont{J.}~\bibnamefont{Dalibard}},
  \bibinfo{journal}{e-print: arXiv/1212.3552}  (\bibinfo{year}{2013}).

\bibitem[{\citenamefont{Walther et~al.}(2006)\citenamefont{Walther, Varcoe,
  Englert, and Becker}}]{Walther-2006}
\bibinfo{author}{\bibfnamefont{H.}~\bibnamefont{Walther}},
  \bibinfo{author}{\bibfnamefont{B.~T.~H.} \bibnamefont{Varcoe}},
  \bibinfo{author}{\bibfnamefont{B.-G.} \bibnamefont{Englert}},
  \bibnamefont{and} \bibinfo{author}{\bibfnamefont{T.}~\bibnamefont{Becker}},
  \bibinfo{journal}{Rep. Prog. Phys.} \textbf{\bibinfo{volume}{69}},
  \bibinfo{pages}{1325} (\bibinfo{year}{2006}).

\bibitem[{\citenamefont{Raimond et~al.}(2001)\citenamefont{Raimond, Brune, and
  Haroche}}]{Raimond-2008}
\bibinfo{author}{\bibfnamefont{J.~M.} \bibnamefont{Raimond}},
  \bibinfo{author}{\bibfnamefont{M.}~\bibnamefont{Brune}}, \bibnamefont{and}
  \bibinfo{author}{\bibfnamefont{S.}~\bibnamefont{Haroche}},
  \bibinfo{journal}{Rev. Mod. Phys.} \textbf{\bibinfo{volume}{73}},
  \bibinfo{pages}{565} (\bibinfo{year}{2001}).

\bibitem[{\citenamefont{Colombe et~al.}(2007)\citenamefont{Colombe, Steinmetz,
  Dubois, Linke, Hunger, and Reichel}}]{Colombe-2007}
\bibinfo{author}{\bibfnamefont{Y.}~\bibnamefont{Colombe}},
  \bibinfo{author}{\bibfnamefont{T.}~\bibnamefont{Steinmetz}},
  \bibinfo{author}{\bibfnamefont{G.}~\bibnamefont{Dubois}},
  \bibinfo{author}{\bibfnamefont{F.}~\bibnamefont{Linke}},
  \bibinfo{author}{\bibfnamefont{D.}~\bibnamefont{Hunger}}, \bibnamefont{and}
  \bibinfo{author}{\bibfnamefont{J.}~\bibnamefont{Reichel}},
  \bibinfo{journal}{Nature} \textbf{\bibinfo{volume}{450}},
  \bibinfo{pages}{272} (\bibinfo{year}{2007}).

\bibitem[{\citenamefont{Ritsch et~al.}(2013)\citenamefont{Ritsch, Domokos,
  Brennecke, and Esslinger}}]{Ritsch-2012}
\bibinfo{author}{\bibfnamefont{H.}~\bibnamefont{Ritsch}},
  \bibinfo{author}{\bibfnamefont{P.}~\bibnamefont{Domokos}},
  \bibinfo{author}{\bibfnamefont{F.}~\bibnamefont{Brennecke}},
  \bibnamefont{and}
  \bibinfo{author}{\bibfnamefont{T.}~\bibnamefont{Esslinger}},
  \bibinfo{journal}{Rev. Mod. Phys.} \textbf{\bibinfo{volume}{85}},
  \bibinfo{pages}{553} (\bibinfo{year}{2013}).

\bibitem[{\citenamefont{Baumann et~al.}(2010)\citenamefont{Baumann, Guerlin,
  Brennecke, and Esslinger}}]{Baumann-2010}
\bibinfo{author}{\bibfnamefont{K.}~\bibnamefont{Baumann}},
  \bibinfo{author}{\bibfnamefont{C.}~\bibnamefont{Guerlin}},
  \bibinfo{author}{\bibfnamefont{F.}~\bibnamefont{Brennecke}},
  \bibnamefont{and}
  \bibinfo{author}{\bibfnamefont{T.}~\bibnamefont{Esslinger}},
  \bibinfo{journal}{Nature} \textbf{\bibinfo{volume}{464}},
  \bibinfo{pages}{1301} (\bibinfo{year}{2010}).

\bibitem[{\citenamefont{Kruse et~al.}(2003)\citenamefont{Kruse, von Cube,
  Zimmermann, and Courteille}}]{Kruse-2003}
\bibinfo{author}{\bibfnamefont{D.}~\bibnamefont{Kruse}},
  \bibinfo{author}{\bibfnamefont{C.}~\bibnamefont{von Cube}},
  \bibinfo{author}{\bibfnamefont{C.}~\bibnamefont{Zimmermann}},
  \bibnamefont{and} \bibinfo{author}{\bibfnamefont{P.~W.}
  \bibnamefont{Courteille}}, \bibinfo{journal}{Phys. Rev. Lett.}
  \textbf{\bibinfo{volume}{91}}, \bibinfo{pages}{183601}
  (\bibinfo{year}{2003}).

\bibitem[{\citenamefont{von Cube et~al.}(2004)\citenamefont{von Cube, Slama,
  Kruse, Zimmermann, Courteille, Robb, Piovella, and Bonifacio}}]{Cube-2004}
\bibinfo{author}{\bibfnamefont{C.}~\bibnamefont{von Cube}},
  \bibinfo{author}{\bibfnamefont{S.}~\bibnamefont{Slama}},
  \bibinfo{author}{\bibfnamefont{D.}~\bibnamefont{Kruse}},
  \bibinfo{author}{\bibfnamefont{C.}~\bibnamefont{Zimmermann}},
  \bibinfo{author}{\bibfnamefont{P.~W.} \bibnamefont{Courteille}},
  \bibinfo{author}{\bibfnamefont{G.~R.~M.} \bibnamefont{Robb}},
  \bibinfo{author}{\bibfnamefont{N.}~\bibnamefont{Piovella}}, \bibnamefont{and}
  \bibinfo{author}{\bibfnamefont{R.}~\bibnamefont{Bonifacio}},
  \bibinfo{journal}{Phys. Rev. Lett.} \textbf{\bibinfo{volume}{93}},
  \bibinfo{pages}{083601} (\bibinfo{year}{2004}).

\bibitem[{\citenamefont{Slama et~al.}(2007{\natexlab{a}})\citenamefont{Slama,
  Bux, Krenz, Zimmermann, and Courteille}}]{Slama-2007}
\bibinfo{author}{\bibfnamefont{S.}~\bibnamefont{Slama}},
  \bibinfo{author}{\bibfnamefont{S.}~\bibnamefont{Bux}},
  \bibinfo{author}{\bibfnamefont{G.}~\bibnamefont{Krenz}},
  \bibinfo{author}{\bibfnamefont{C.}~\bibnamefont{Zimmermann}},
  \bibnamefont{and} \bibinfo{author}{\bibfnamefont{P.~W.}
  \bibnamefont{Courteille}}, \bibinfo{journal}{Phys. Rev. Lett.}
  \textbf{\bibinfo{volume}{98}}, \bibinfo{pages}{053603}
  (\bibinfo{year}{2007}{\natexlab{a}}).

\bibitem[{\citenamefont{Slama et~al.}(2007{\natexlab{b}})\citenamefont{Slama,
  Krenz, Bux, Zimmermann, and Courteille}}]{Slama-2007-A}
\bibinfo{author}{\bibfnamefont{S.}~\bibnamefont{Slama}},
  \bibinfo{author}{\bibfnamefont{G.}~\bibnamefont{Krenz}},
  \bibinfo{author}{\bibfnamefont{S.}~\bibnamefont{Bux}},
  \bibinfo{author}{\bibfnamefont{C.}~\bibnamefont{Zimmermann}},
  \bibnamefont{and} \bibinfo{author}{\bibfnamefont{P.~W.}
  \bibnamefont{Courteille}}, \bibinfo{journal}{Phys. Rev. A}
  \textbf{\bibinfo{volume}{75}}, \bibinfo{pages}{063620}
  (\bibinfo{year}{2007}{\natexlab{b}}).

\bibitem[{\citenamefont{Brattke et~al.}(2001)\citenamefont{Brattke, Varcoe, and
  Walther}}]{Brattke-2001}
\bibinfo{author}{\bibfnamefont{S.}~\bibnamefont{Brattke}},
  \bibinfo{author}{\bibfnamefont{B.~T.~H.} \bibnamefont{Varcoe}},
  \bibnamefont{and} \bibinfo{author}{\bibfnamefont{H.}~\bibnamefont{Walther}},
  \bibinfo{journal}{Phys. Rev. Lett.} \textbf{\bibinfo{volume}{86}},
  \bibinfo{pages}{3534} (\bibinfo{year}{2001}).

\bibitem[{\citenamefont{McKeever et~al.}(2004)\citenamefont{McKeever, Boca,
  Boozer, Miller, Buck, Kuzmich, and Kimble}}]{McKeever-2004}
\bibinfo{author}{\bibfnamefont{J.}~\bibnamefont{McKeever}},
  \bibinfo{author}{\bibfnamefont{A.}~\bibnamefont{Boca}},
  \bibinfo{author}{\bibfnamefont{A.~D.} \bibnamefont{Boozer}},
  \bibinfo{author}{\bibfnamefont{R.}~\bibnamefont{Miller}},
  \bibinfo{author}{\bibfnamefont{J.~R.} \bibnamefont{Buck}},
  \bibinfo{author}{\bibfnamefont{A.}~\bibnamefont{Kuzmich}}, \bibnamefont{and}
  \bibinfo{author}{\bibfnamefont{H.~J.} \bibnamefont{Kimble}},
  \bibinfo{journal}{Science} \textbf{\bibinfo{volume}{33}},
  \bibinfo{pages}{1992} (\bibinfo{year}{2004}).

\bibitem[{\citenamefont{Keller et~al.}(2004)\citenamefont{Keller, Lange,
  Hayasaka, Lange, and Walther}}]{Keller-2004}
\bibinfo{author}{\bibfnamefont{M.}~\bibnamefont{Keller}},
  \bibinfo{author}{\bibfnamefont{B.}~\bibnamefont{Lange}},
  \bibinfo{author}{\bibfnamefont{K.}~\bibnamefont{Hayasaka}},
  \bibinfo{author}{\bibfnamefont{W.}~\bibnamefont{Lange}}, \bibnamefont{and}
  \bibinfo{author}{\bibfnamefont{H.}~\bibnamefont{Walther}},
  \bibinfo{journal}{Nature} \textbf{\bibinfo{volume}{431}},
  \bibinfo{pages}{1075} (\bibinfo{year}{2004}).

\bibitem[{\citenamefont{Cooper et~al.}(2013)\citenamefont{Cooper, Wright,
  S\"{o}ller, and Smith}}]{Merlin-Cooper-2013}
\bibinfo{author}{\bibfnamefont{M.}~\bibnamefont{Cooper}},
  \bibinfo{author}{\bibfnamefont{L.~J.} \bibnamefont{Wright}},
  \bibinfo{author}{\bibfnamefont{C.}~\bibnamefont{S\"{o}ller}},
  \bibnamefont{and} \bibinfo{author}{\bibfnamefont{B.~J.} \bibnamefont{Smith}},
  \bibinfo{journal}{Opt. Express} \textbf{\bibinfo{volume}{21}},
  \bibinfo{pages}{5309} (\bibinfo{year}{2013}).

\bibitem[{\citenamefont{Gerry and Eberly}(1990)}]{Gerry-1990}
\bibinfo{author}{\bibfnamefont{C.~C.} \bibnamefont{Gerry}} \bibnamefont{and}
  \bibinfo{author}{\bibfnamefont{J.~H.} \bibnamefont{Eberly}},
  \bibinfo{journal}{Phys. Rev. A} \textbf{\bibinfo{volume}{42}},
  \bibinfo{pages}{6805} (\bibinfo{year}{1990}).

\bibitem[{\citenamefont{Shore and Knight}(1993)}]{Shore-1993}
\bibinfo{author}{\bibfnamefont{B.}~\bibnamefont{Shore}} \bibnamefont{and}
  \bibinfo{author}{\bibfnamefont{P.}~\bibnamefont{Knight}},
  \bibinfo{journal}{J. Mod. Opt.} \textbf{\bibinfo{volume}{40}},
  \bibinfo{pages}{1195} (\bibinfo{year}{1993}).

\bibitem[{\citenamefont{Biedenharn and Dam}(1965)}]{Schwinger-1965}
\bibinfo{editor}{\bibfnamefont{L.~C.} \bibnamefont{Biedenharn}}
  \bibnamefont{and} \bibinfo{editor}{\bibfnamefont{H.~V.} \bibnamefont{Dam}},
  eds., \emph{\bibinfo{title}{Quantum Theory of Angular Momentum}}
  (\bibinfo{publisher}{Academic Press}, \bibinfo{year}{1965}).

\bibitem[{\citenamefont{Angelakis et~al.}(2007)\citenamefont{Angelakis, Santos,
  , and Bose}}]{Angelakis-2007}
\bibinfo{author}{\bibfnamefont{D.~G.} \bibnamefont{Angelakis}},
  \bibinfo{author}{\bibfnamefont{M.~F.} \bibnamefont{Santos}}, ,
  \bibnamefont{and} \bibinfo{author}{\bibfnamefont{S.}~\bibnamefont{Bose}},
  \bibinfo{journal}{Phys. Rev. A} \textbf{\bibinfo{volume}{76}},
  \bibinfo{pages}{031805} (\bibinfo{year}{2007}).

\bibitem[{\citenamefont{Koch and Hur}(2009)}]{Koch-2009}
\bibinfo{author}{\bibfnamefont{J.}~\bibnamefont{Koch}} \bibnamefont{and}
  \bibinfo{author}{\bibfnamefont{K.~L.} \bibnamefont{Hur}},
  \bibinfo{journal}{Phys. Rev. A} \textbf{\bibinfo{volume}{80}},
  \bibinfo{pages}{023811} (\bibinfo{year}{2009}).

\bibitem[{\citenamefont{Brahms et~al.}(2011)\citenamefont{Brahms, Purdy,
  Brooks, Botter, and Stamper-Kurn}}]{Brahms-2011}
\bibinfo{author}{\bibfnamefont{N.}~\bibnamefont{Brahms}},
  \bibinfo{author}{\bibfnamefont{T.~P.} \bibnamefont{Purdy}},
  \bibinfo{author}{\bibfnamefont{D.~W.~C.} \bibnamefont{Brooks}},
  \bibinfo{author}{\bibfnamefont{T.}~\bibnamefont{Botter}}, \bibnamefont{and}
  \bibinfo{author}{\bibfnamefont{D.~M.} \bibnamefont{Stamper-Kurn}},
  \bibinfo{journal}{Nat. Phys.} \textbf{\bibinfo{volume}{7}},
  \bibinfo{pages}{604} (\bibinfo{year}{2011}).

\bibitem[{\citenamefont{Fisher and Fradkin}(1985)}]{Fisher-1985}
\bibinfo{author}{\bibfnamefont{M.~P.~A.} \bibnamefont{Fisher}}
  \bibnamefont{and} \bibinfo{author}{\bibfnamefont{E.}~\bibnamefont{Fradkin}},
  \bibinfo{journal}{Nuclear Physics B} \textbf{\bibinfo{volume}{251}},
  \bibinfo{pages}{457} (\bibinfo{year}{1985}).

\bibitem[{\citenamefont{Sebastian et~al.}(2012)\citenamefont{Sebastian,
  Harrison, Liang, Bonn, Hardy, Mielke, and Lonzarich}}]{Sebastian-2012}
\bibinfo{author}{\bibfnamefont{S.~E.} \bibnamefont{Sebastian}},
  \bibinfo{author}{\bibfnamefont{N.}~\bibnamefont{Harrison}},
  \bibinfo{author}{\bibfnamefont{R.}~\bibnamefont{Liang}},
  \bibinfo{author}{\bibfnamefont{D.~A.} \bibnamefont{Bonn}},
  \bibinfo{author}{\bibfnamefont{W.~N.} \bibnamefont{Hardy}},
  \bibinfo{author}{\bibfnamefont{C.~H.} \bibnamefont{Mielke}},
  \bibnamefont{and} \bibinfo{author}{\bibfnamefont{G.~G.}
  \bibnamefont{Lonzarich}}, \bibinfo{journal}{Phys. Rev. Lett.}
  \textbf{\bibinfo{volume}{108}}, \bibinfo{pages}{196403}
  (\bibinfo{year}{2012}).

\bibitem[{\citenamefont{Spielman}(2009)}]{Spielman-2009}
\bibinfo{author}{\bibfnamefont{I.~B.} \bibnamefont{Spielman}},
  \bibinfo{journal}{Phys. Rev. A} \textbf{\bibinfo{volume}{79}},
  \bibinfo{pages}{063613} (\bibinfo{year}{2009}).

\bibitem[{\citenamefont{Parkins et~al.}(1993)\citenamefont{Parkins, Marte,
  Zoller, and Kimble}}]{Parkins-1993}
\bibinfo{author}{\bibfnamefont{A.~S.} \bibnamefont{Parkins}},
  \bibinfo{author}{\bibfnamefont{P.}~\bibnamefont{Marte}},
  \bibinfo{author}{\bibfnamefont{P.}~\bibnamefont{Zoller}}, \bibnamefont{and}
  \bibinfo{author}{\bibfnamefont{H.~J.} \bibnamefont{Kimble}},
  \bibinfo{journal}{Phys. Rev. Lett.} \textbf{\bibinfo{volume}{71}},
  \bibinfo{pages}{3095} (\bibinfo{year}{1993}).

\bibitem[{\citenamefont{Parkins et~al.}(1995)\citenamefont{Parkins, Marte,
  Zoller, Carnal, and Kimble}}]{Parkins-1995}
\bibinfo{author}{\bibfnamefont{A.~S.} \bibnamefont{Parkins}},
  \bibinfo{author}{\bibfnamefont{P.}~\bibnamefont{Marte}},
  \bibinfo{author}{\bibfnamefont{P.}~\bibnamefont{Zoller}},
  \bibinfo{author}{\bibfnamefont{O.}~\bibnamefont{Carnal}}, \bibnamefont{and}
  \bibinfo{author}{\bibfnamefont{H.~J.} \bibnamefont{Kimble}},
  \bibinfo{journal}{Phys. Rev. A} \textbf{\bibinfo{volume}{51}},
  \bibinfo{pages}{1578} (\bibinfo{year}{1995}).

\bibitem[{\citenamefont{Gogyan et~al.}(2012)\citenamefont{Gogyan, Gu\'{e}rin,
  Leroy, and Malakyan}}]{Gogyan-2012}
\bibinfo{author}{\bibfnamefont{A.}~\bibnamefont{Gogyan}},
  \bibinfo{author}{\bibfnamefont{S.}~\bibnamefont{Gu\'{e}rin}},
  \bibinfo{author}{\bibfnamefont{C.}~\bibnamefont{Leroy}}, \bibnamefont{and}
  \bibinfo{author}{\bibfnamefont{Y.}~\bibnamefont{Malakyan}},
  \bibinfo{journal}{Phys. Rev. A} \textbf{\bibinfo{volume}{86}},
  \bibinfo{pages}{063801} (\bibinfo{year}{2012}).

\bibitem[{\citenamefont{Scully and Zubairy}(1997)}]{Scully-1997}
\bibinfo{author}{\bibfnamefont{M.}~\bibnamefont{Scully}} \bibnamefont{and}
  \bibinfo{author}{\bibfnamefont{M.~S.} \bibnamefont{Zubairy}},
  \emph{\bibinfo{title}{Quantum Optics}} (\bibinfo{publisher}{Cambridge
  University Press}, \bibinfo{year}{1997}).

\bibitem[{\citenamefont{Meystre and Sargent}(1999)}]{Meystre1999}
\bibinfo{author}{\bibfnamefont{P.}~\bibnamefont{Meystre}} \bibnamefont{and}
  \bibinfo{author}{\bibfnamefont{M.}~\bibnamefont{Sargent}},
  \emph{\bibinfo{title}{Elements of Quantum Optics, 3rd. ed.}}
  (\bibinfo{publisher}{Springer}, \bibinfo{year}{1999}).

\bibitem[{\citenamefont{Kaplan}(1983)}]{Kaplan-1983}
\bibinfo{author}{\bibfnamefont{T.}~\bibnamefont{Kaplan}}, \bibinfo{journal}{Z.
  Phys. B} \textbf{\bibinfo{volume}{49}}, \bibinfo{pages}{313}
  (\bibinfo{year}{1983}).

\bibitem[{\citenamefont{Maschler et~al.}(2008)\citenamefont{Maschler, Mekhov,
  and Ritsch}}]{Maschler-2008}
\bibinfo{author}{\bibfnamefont{C.}~\bibnamefont{Maschler}},
  \bibinfo{author}{\bibfnamefont{I.~B.} \bibnamefont{Mekhov}},
  \bibnamefont{and} \bibinfo{author}{\bibfnamefont{H.}~\bibnamefont{Ritsch}},
  \bibinfo{journal}{Eur. Phys. J. D} \textbf{\bibinfo{volume}{46}},
  \bibinfo{pages}{545} (\bibinfo{year}{2008}).

\bibitem[{\citenamefont{M\"{u}ller et~al.}(2012)\citenamefont{M\"{u}ller,
  Strack, and Sachdev}}]{Muller-2012}
\bibinfo{author}{\bibfnamefont{M.}~\bibnamefont{M\"{u}ller}},
  \bibinfo{author}{\bibfnamefont{P.}~\bibnamefont{Strack}}, \bibnamefont{and}
  \bibinfo{author}{\bibfnamefont{S.}~\bibnamefont{Sachdev}},
  \bibinfo{journal}{Phys. Rev. A} \textbf{\bibinfo{volume}{86}},
  \bibinfo{pages}{023604} (\bibinfo{year}{2012}).

\end{thebibliography}
 
\end{document}